\documentclass[%
 reprint,
superscriptaddress,
bibnotes,
 amsmath,amssymb,
 aps,
]{revtex4-1}

\usepackage{graphicx}% Include figure files
\usepackage{dcolumn}% Align table columns on decimal point
\usepackage{bm}% bold math
\usepackage[T1]{fontenc}
\usepackage{mathptmx}
\usepackage{mathrsfs}
\usepackage{xcolor}
\usepackage{hyperref}
\usepackage{physics}
\usepackage{simplewick}
\usepackage{mathrsfs}
\usepackage[bottom]{footmisc}

\newcommand{\be}{\begin{equation}}
\newcommand{\ee}{\end{equation}}
\newcommand{\bs}{\begin{split}}
\newcommand{\es}{\end{split}}

\begin{document}
\title{Linear quantum measurement theory of matter-wave interferometry}

\author{Yiqiu Ma}
\email{myqphy@gmail.com}
\affiliation{Center for Gravitational Experiment, Ministry of Education key Laboratory for fundamental physical quantity measurement, School of Physics, Huazhong University of Science and Technology, Wuhan, 430074, China}
\affiliation{Theoretical Astrophysics 350-17, California Institute of Technology, Pasadena, California 91125, USA}

\author{Xiang Li}
\affiliation{Theoretical Astrophysics 350-17, California Institute of Technology, Pasadena, California 91125, USA}

\author{Shengjun Yang}
\affiliation{College of Science, Southern University of Science and Technology, Shenzhen,518055, China}

\author{Yanbei Chen}
\email{yanbei@caltech.edu}
\affiliation{Theoretical Astrophysics 350-17, California Institute of Technology, Pasadena, California 91125, USA}
\date{\today}% It is always \today, today,

\begin{abstract}
The theory of linear quantum measurement has been developed for analysing the
sensitivities of experimental devices that measure extremely
weak signals, such as gravitational waves. It has successfully contributed to
the theoretical understanding of laser interferometer gravitational-wave detectors (used by LIGO, VIRGO and KaGRA) and helped many important experimental upgrades. In this work, we establish a linear quantum measurement theory for another kind of measurement device--- matter wave interferometers, which has been widely discussed as an important platform for many high-precision experiments. This theory allows us to account for both atom and light fluctuations, and leads to a detailed analysis of back-action in matter-wave interferometry (action of light back onto the atoms) and its effect on dynamics and measurement noise. From this analysis, we obtain a Standard Quantum Limit (SQL) for matter-wave interferometry.  A comparison between the LIGO detector and matter wave interferometer is also given from the perspective of quantum measurement. 
\end{abstract}

\maketitle

\section{Introduction}

The detection of gravitational waves (GWs) from merging binary black holes\,\cite{GW150914} and merging neutron star binaries\,\cite{GW170817} by an international network of gravitational-wave detectors (LIGO, VIRGO and KAGRA) opened the era of gravitational wave astronomy (in this paper we shall use "LIGO detector" to refer to a detector that is used by LIGO\,\cite{ligo2010}, VIRGO\,\cite{virgo2015} and KaGRA\,\cite{kagra2019}). This detection is also a milestone in the development of high-precision measurement physics, making LIGO detector the most sensitive instrument that human beings ever built. Parallel to LIGO detector where the underline principle is the interference of the electromagnetic waves, other concepts of GW detectors have also been proposed, even before the first detection event. One particular attractive concept is the atom-interferometer GW detector, first raised by Dimopoulos {\it et. al}\,\cite{Dimopoulos2008,Dimopoulos2008prd} and later enriched by many further discussions\,\cite{Tino2007,Yu2011,Bender2012,Graham2013,Harms2013,Geiger2016,Graham2018,Junca2019,Zhan2019}. Different from the LIGO, the physical principles under the atom interferometer GW detector is the interference of the matter waves, rather than the light waves.

The concept of atom interferometer can be traced back to the 1930s, when Rabi demonstrated that the atoms' internal quantum states can be altered using rf resonance\,\cite{Rabi1938}. In 1949, Ramsey firstly created and detected long-lived coherent superposition of internal quantum states\,\cite{Ramsey1950}. These pioneering works pave the way for the further development of a field named \emph{atom optics}, namely, one can manipulate coherent beams of atoms as manipulating that of light fields\,\cite{Cronin2009}. Atom interferometry is an art of atom optics and an important experimental platform for high-precision measurement, which is now being used for measuring earth's gravity acceleration and testing fundamental physics\,\cite{Dimopoulos2007,Mueller2008,Arvanitaki2008,Dimopoulos2008prd2,Chung2009,Bouchendira2011,Hohensee2011,Hu2013,Rosi2014,Hamilton2015,Duan2016,Elder2016,Parker2018}.

The advantage of the proposed application of atom interferometer in GW detection is mostly at low frequency (below 10 Hz), which can be understood as follows. Because the test masses are connected to the ground through suspension system, the sensitivity of a laser interferometer GW detector is seriously contaminated at low frequencies partly through the coupling of the test masses with the sesmic oscillations, although the multi-stage vibration isolation technique has been applied\,\cite{Harms2013}. For space-borne optical GW detector such as Laser Interferometer Space Antenna (LISA), the test masses are also connected to the satellite platforms thereby the random motion of the satellites will be transfered onto the test masses and contaminate the GW signal. However, for the atom interferometer, since the atoms are free-falling during the interferometry process, they are less sensitive to the seismic  perturbation (or the satellite motion in the space case). The laser noise can be removed by designing the detector configuration with common mode rejection. More sophisticated designs such as implementing the large momentum transfer technique or optical cavities have been also discussed\,\cite{Mueller2008lmt,Clade2009,Onur2016,Hamilton2015cavity, Canuel2018}.

\begin{figure}[h]
   \centering
   \includegraphics[scale=0.10]{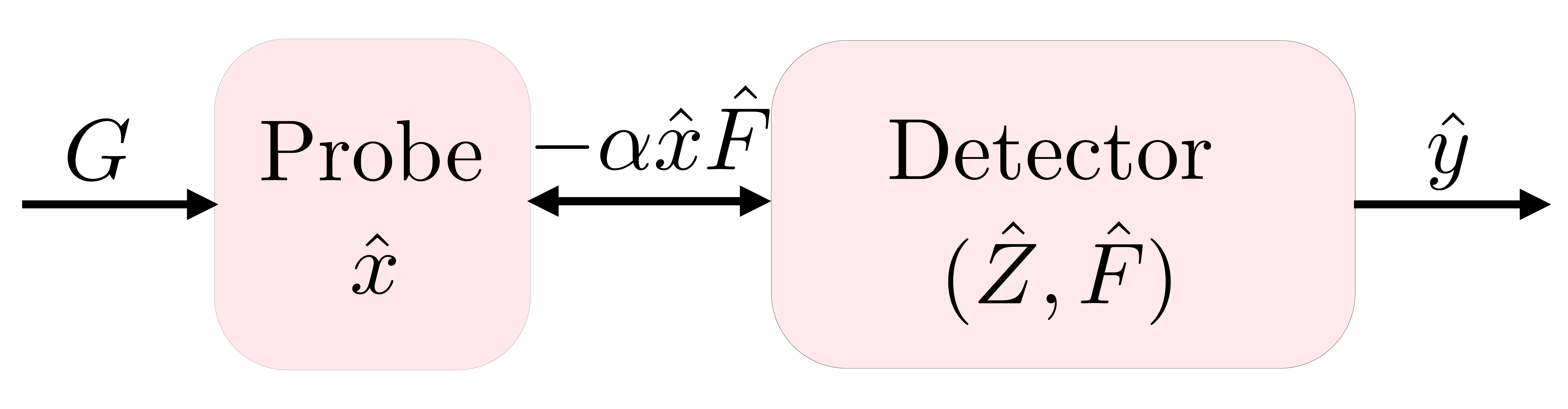} % requires the graphicx package
   \caption{Block diagram of a linear quantum measurement device.}
   \label{fig:quantum_measurement}
\end{figure}

Typically, experimental devices such as GW detectors that targeted on measuring extremely weak signals can be even affected by the quantum mechanics. The theory of quantum measurement developed from 1960s is a framework to analyse how quantum mechanics affects the sensitivity of an experimental device\,\cite{Braginsky1992}.  The early resonant bar GW detectors and the current laser interferometer GW detectors have been extensively studied and understood using this quantum measurement theory framework\,\cite{Braginsky1992, Kimble2001, Buonanno2002}. For atom interferometry, although the effect of quantum noise has been discussed by various authors\,\cite{Itano1993,Jacobson1995,Dowling1998,Scully1993}, a complete analysis under quantum measurement theory has not been discussed in the current literatures. Establishing such a theory will provide important insights in understanding the atom interferometer. Here, it is useful to briefly overview such a framework, based on the block-diagram shown in Fig.\,\ref{fig:quantum_measurement}.
In this framework, a quantum measurement device is divided into probe and detector, where the probe dynamical quantity $\hat x$ is linearly coupled to $G$---the information to be measured. The probe and detector are coupled through linear Hamiltonian $\hat H_{\rm int}=-\alpha\hat x\hat F$. The information of $G$ will flow into the detector through probe-detector interaction and then be read out as $\hat y(t)$:
\be\label{eq:qmreadout}
\hat y(t)=\hat Z(t)+\alpha\left[\hat x_{\rm zero}(t)+x_{\rm sig}(t)+\alpha\int^t_{-\infty}dt'\chi_{m}(t-t')\hat F(t')\right],
\ee
where $\hat x_{\rm zero}$, $x_{\rm sig}$, $\hat F$ are the zero-point fluctuation of $\hat x$, the signal, and the back-action force, $\alpha$ is the coupling strength and the $\chi_{m}(t-t')$ is the dynamical response function of the probe. Braginsky {\it et.al} shows that for measuring the GW tidal force, the zero-point fluctuation of test masses does not contribute to the final sensitivity thereby $\hat x_{\rm zero}$ can be simply ignored\,\cite{Braginsky2003}. If $\hat Z$ and $\hat F$ have no correlation, the sensitivity will be limited by the so-called standard quantum limit (SQL), given by (in the frequency domain):
\be\label{eq:sql}
S^{\rm SQL}_{xx}(\Omega)=2\hbar|\chi_{m}(\Omega)|,
\ee
where $\Omega$ is the angular frequency of GWs. For advanced LIGO, the SQL is given by $S^{\rm SQL}_{xx}(\Omega)=2\hbar/(m\Omega^2)$, with the mass of the test mirrors denoted by $m$.

In this work, we set up a quantum measurement theory framework for analysing the physics of atom interferometer, which is based on the interaction between atom cloud and two optical fields (passive and control laser). It is straightforward to extend our result to other atom interferometer configurations.

\section{Effective Hamiltonian of an atom interferometer}
In an atom interferometer GW detector, the GW information is carried by the light field in the TT gauge, thereby the light field corresponds to the probe  and the detector corresponds to the atom cloud in the above quantum measurement model. Concretely speaking, the atom cloud, as a phase meter, records the optical phase (more precisely, the phase difference between the control and passive fields as we shall see) imposed by the signal. In this section, we will establish an one-dimensional effective Hamiltonian for analysing this system. Real systems are three-dimensional therefore this one-dimensional model is obtained by reducing a three-dimensional system by paraxial approximation. This Hamiltonian can be derived from first principle and the details are given in Appendix. In this section, we are going to show how the back-action effect manifests itself in atom interferometers.

%and after the interaction, the light fields propagate out of the atom-light interaction region will carry the information of fluctuating atoms. In a single atom interferometer, the two reflection processes are connected by the same control field and the control field flies out of the first reflection region will carrying atom information and impose a back action on the atoms of the second reflection process. Similarly, in an atom-interferometric gravitational wave detector, a typical configuration is two atomic interferometers, linked by the same control light pulse in each step. The signal is embedded in the phase difference between two interferometers. Possible quantum back-action lies on the transfer of quantum noise from the first interferometer to control pulse and then from the control pulse to the second one. If it can influence the dynamics of atom-light interaction in the second interferometer, the noise in two outputs will be entangled and this will give rise to the ultimate limit. In this section, we will describe an effective Hamiltonian for analysing the atom interferometer GW detector, which can be more derived from first principle using field theory language (see the Appendix). 

\subsection{Effective Hamiltonian and dynamics of an interaction kernel}
\begin{figure}[h]
   \centering
   \includegraphics[scale=0.65]{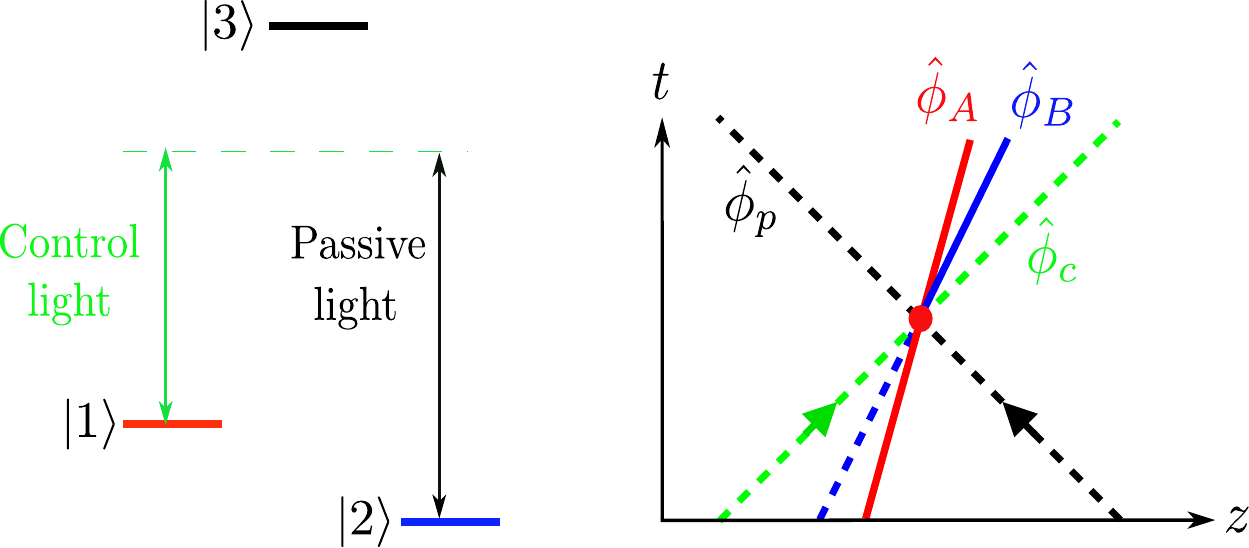} % requires the graphicx package
   \caption{A four-boson Raman interaction kernel and its corresponding WKB trajectory. For a detailed discussion of this process and the description using field theory, see the Appendix.}
   \label{fig:interaction}
\end{figure}

The basic physical process happen in a typical atom interferometer is a four-boson interaction, where the atomic transition between energy level $|1\rangle$ and $|2\rangle$ happens through  coupling to an intermediate energy level $|3\rangle$ by control and passive fields, as shown in Fig\,\ref{fig:interaction}.

The Hamiltonian describing the Raman interaction happens in an atom interferometer  has the following structure:
\be\label{eq:strawmanH}
\begin{split}
&\hat H_{\rm opt}=\frac{i\hbar}{2}\int^\infty_{-\infty}[\partial_x\hat a_{cx}^\dag\hat a_{cx}-\partial_x\hat a_{px}^\dag\hat a_{px}]+{\rm h.c},\\
&\hat H_{\rm a}= \hbar\omega_A\hat A^\dag\hat A+\hbar\omega_B\hat B^\dag\hat B,\\
&H_{\rm int}= \hbar \chi(\hat A+\hat A^\dag)(\hat B+\hat B^\dag)(\hat a_c+\hat a_c^\dag)(\hat a_p+\hat a_p^\dag),
\end{split}
\ee
where the $H_{\rm opt}$ describes the free control light $\hat a_c$ and passive light $\hat a_p$ in $x-$space; $\hat H_{\rm a}$ describes the whole atom clouds at two different energy levels and thereby does not depend on $x$; $\hat H_{\rm int}$ describes the atom-light Raman interaction at one specific spacetime location, the derivation of such a four-field interaction Hamiltonian is shown in the Appendix. The $\hat A$ and $\hat B$ are the effective annihilation operators for the energy level $|1\rangle$ and $|2\rangle$, respectively. Their corresponding number operators are $\hat N_A=\hat A^\dag\hat A$ and $\hat N_B=\hat B^\dag\hat B$ and we have the commutation relation $[\hat A,\hat A^\dag]=1$ (the same for $\hat B$). For the continuous optical fields, we have $[\hat a_{cx},\hat a^\dag_{cx'}]=\delta(x-x')$ (the same for $\hat a_{px}$) and it is related to particle numeber by $\hat N=\int dx\hat a_{cx}^\dag\hat a_{cx}$. The precise definition of these effective operators will be presented in Section IV. 

Note that under the rotating wave approximations, only the terms satisfying $\omega_A+\omega_c=\omega_B+\omega_p$ will be kept while those non-rotating wave terms such as $\hat A\hat B\hat a_c^\dag\hat a_p^\dag$\,\ \emph{etc.} can be safely ignored, which leads us to a simpler form of interaction Hamiltonian:
\be\label{eq: interactionH}
\hat H_{\rm int}=\hbar \chi\hat A^\dag\hat B\hat a_c^\dag\hat a_p+{\rm h.c},
\ee 
and the corresponding equations of motion for atomic clouds are given by:
\be
\dot{\hat A}=-i\chi \hat B\hat a_c^\dag\hat a_p,\quad\dot{\hat B}=-i\chi\hat A\hat a_c\hat a_p^\dag.
\ee

Solving these equations perturbatively by writing the operators as: $\hat A=\bar A_A+ \hat A_A$ and $\hat B=\bar A_B+\hat A_B$ (where the magnitude of $\hat A_{A/B}$ is small compare to the $\bar A_{A/B}$) leads to the zeroth-order equations:
\be
\dot{\bar A}_A=-i\Omega\bar A_B e^{i\varphi},\quad\dot{\bar A}_B=-i\Omega\bar A_A e^{-i\varphi}.
\ee
where $\Omega=\chi |\bar a_c\bar a_p| $  is the Rabi-frequency and $\varphi=\varphi_c-\varphi_p$ which is the phase difference between control field and passive field.

Then the rest terms satisfy the following equations:
\be\label{eq:atomeom}
\begin{split}
\mathcal{\hat L}
\begin{bmatrix}
\hat A_A\\
\hat A_B
\end{bmatrix}
=&-i\chi
\begin{bmatrix}
\bar A_B(\hat a_c^\dag|\bar a_p|e^{-i\varphi_p}+\hat a_p |\bar a_c|e^{i\varphi_c})\\
\bar A_A(\hat a_c |\bar a_p|e^{i\varphi_p}+\hat a_p^\dag |\bar a_c|e^{-i\varphi_c})
\end{bmatrix}\\
&+\Omega\varphi_s
\begin{bmatrix}
\bar A_B e^{i\varphi}\\
-\bar A_A e^{-i\varphi}
\end{bmatrix}.
\end{split}
\ee
%Then the rest terms satisfy the following equations:
%\be\label{eq:atomeom}
%\begin{split}
%&\dot{\hat A}_A+i\Omega e^{i\varphi}\hat A_B=-i\chi\bar A_B(\hat a_c^\dag|\bar a_p|e^{-i\varphi_p}+\hat a_p |\bar a_c|e^{i\varphi_c}),\\
%&\dot{\hat A}_B+i\Omega e^{-i\varphi}\hat A_A=-i\chi\bar A_A(\hat a_c |\bar a_p|e^{i\varphi_p}+\hat a_p^\dag |\bar a_c|e^{-i\varphi_c}),\\
%\end{split}
%\ee
in which the differential operator $\mathcal{\hat L}$ takes the form of:
\be
\mathcal{\hat L}=
\begin{bmatrix}
\partial_t&i\Omega e^{i\varphi}\\
i\Omega e^{-i\varphi}&\partial_t
\end{bmatrix},
\ee
and the right hand side of these equations describes the influence of optical fields to the evolution of atom fields.  The $\varphi_s\ll\varphi$ is the signal phase carried by the optical field, and we expand it to the linear order to obtain the signal terms.

The solutions for the optical fields, to the leading order, are given by:
\be\label{eq:opticaleommean}
\bar a_{c\rm out}=\bar a_{c\rm in},\quad \bar a_{p\rm out}=\bar a_{p\rm in}.
\ee
and the rest terms satisfy:
\be\label{eq:opticaleom}
\begin{split}
&\hat a_{c\rm out}=\hat a_{c\rm in}-i\chi(\bar A^*_A\bar A_B\hat a_p+\bar A^*_A\bar a_p\hat A_B+\bar A_B\bar a_p\hat A_A^\dag+\bar A^*_A\bar A_B\bar a_{p}),\\
&\hat a_{p\rm out}=\hat a_{p\rm in}-i\chi(\bar A_A\bar A_B^*\hat a_c+\bar A_A\bar a_c\hat A_B^\dag+\bar A_B^*\bar a_c\hat A_A+\bar A_A\bar A_B^*\bar a_c).\\
\end{split}
\ee

The first terms in the brackets of the r.h.s of the Eq.\,\eqref{eq:opticaleom} are much smaller than the rest terms (the ratio$\sim \sqrt{N_a/N_L}$ where $N_a$ and $N_L$ are the atom number and photon number in the pulse), which can be ignored. Also note that the optical operators on the r.h.s of the above equations are defined at the interaction point, in principle the Eq.\,\ref{eq:atomeom} should be solved in the way that we substitute $\hat a_p=(\hat a_{p\rm in}+\hat a_{p\rm out})/2$ and $\hat a_c=(\hat a_{c\rm in}+\hat a_{c\rm out})/2$. However, since  the atom-light interaction is weak, if we ignore the term involving high orders of $\chi$, the r.h.s. of Eq.\,\eqref{eq:opticaleom} can be simply written in the way that $\hat a_c=\hat a_{c\rm in}$
, $\hat a_p=\hat a_{p\rm in}$.

Substituting Eq.\,\eqref{eq:opticaleom} into the equations of motion for the atom field Eq.\,\eqref{eq:atomeom}, we obtain:
\be\label{eq:fullatomeom}
\mathcal{\hat L}
\begin{bmatrix}
\hat A_A\\
\hat A_B
\end{bmatrix}
=\begin{bmatrix}
\hat F^A_{\rm flu}+F^A_{\rm cl}+F^{A}_{\rm dy}\\
\hat F^B_{\rm flu}+F^B_{\rm cl}+F^B_{\rm dy}
\end{bmatrix}
+\Omega\varphi_s
\begin{bmatrix}
\bar A_B e^{i\varphi}\\
-\bar A_A e^{-i\varphi}
\end{bmatrix}.
\ee
Here the terms on the r.h.s can be understood as "optical force" acting on the atomic fields, explained in detail as follows:

(1) The $\hat F^{A/B}_{\rm flu}$ have the form:
\be\label{eq:langevin_force}
\begin{split}
\hat F^A_{\rm flu}=-i\chi \bar A_B (|\bar a_p|e^{-i\varphi_p}\hat a_{c\rm in}^\dag+|\bar a_c|\hat a_{p\rm in}e^{i\varphi_c}),\\
\hat F^B_{\rm flu}=-i\chi\bar A_A(|\bar a_p|e^{i\varphi_p}\hat a_{c\rm in}+|\bar a_c|e^{-i\varphi_c}\hat a_{p\rm in}^\dag ),
\end{split}
\ee
which is the optical Langevin force acting on the atom fields, due to the randomness of the incoming states of optical fields.

(2) The $F^{A/B}_{\rm cl}$ have the form:
\be\label{eq:mean_force}
\begin{split}
F^A_{\rm cl}=\frac{\chi^2}{2}|\bar A_B|^2\bar A_A(|\bar a_p|^2-|\bar a_c|^2),\\
F^B_{\rm cl}=\frac{\chi^2}{2}|\bar A_A|^2\bar A_B(|\bar a_p|^2-|\bar a_c|^2),
\end{split}
\ee
which describes the pondermotive force exerted by the mean optical fields on the atoms. 

(3) The $F^{A/B}_{\rm dy}$ have the form:
\be\label{eq:dynamical_modification}
\begin{split}
F^A_{\rm dy}=\frac{\chi^2}{2}(|\bar a_p|^2-|\bar a_c|^2)(\bar A_A\bar A_B\hat A_B^\dag+|A_B|^2\hat A_A),\\
F^B_{\rm dy}=\frac{\chi^2}{2}(|\bar a_p|^2-|\bar a_c|^2)(\bar A_B\bar A_A\hat A_A^\dag+|A_A|^2\hat A_B).
\end{split}
\ee
These forces, which depends on the operators $\hat A_A^{(\dag)},\hat A_B^{(\dag)}$ will modify the Rabi-dynamics of the atom fields.

%\be\label{eq:fullatomeom}
%\begin{split}
%&\dot{\hat A}_A+i\Omega e^{i\varphi}\hat A_B=-i\chi \bar A_B (|\bar a_p|e^{-i\varphi_p}\hat a_{c\rm in}^\dag+|\bar a_c|\hat a_{p\rm in}e^{i\varphi_c})\\
%&+\frac{\chi^2}{2}(|\bar a_p|^2-|\bar a_c|^2)(|\bar A_B|^2\bar A_A+\bar A_A\bar A_B\hat A_B^\dag+|A_B|^2\hat A_A),\\
%&\dot{\hat A}_B+i\Omega e^{-i\varphi}\hat A_A=-i\chi\bar A_A(\hat a_c |\bar a_p|e^{i\varphi_p}+\hat a_p^\dag |\bar a_c|e^{-i\varphi_c})\\
%& \frac{\chi^2}{2}(|\bar a_p|^2-|\bar a_c|^2)(|\bar A_A|^2\bar A_B+\bar A_B\bar A_A\hat A_A^\dag+|A_A|^2\hat A_B),
%\end{split}
%\ee
%where the second term on the r.h.s. of Eq.\,\eqref{eq:fullatomeom} can modify the Rabi-dynamics. 
In case of balanced passive and control fields (i.e. $|\bar a_p|=|\bar a_c|=\bar a_L$), these dynamical back-actions will vanish. In the following sections, we will focus on the configuration with balanced passive and control fields.

The solution of these dynamical equations can be expressed in a more convenient way by using the following basis: (1) $\hat A_{\pm}=(\hat A_A e^{-i\varphi/2}\pm\hat A_B e^{i\varphi/2})/\sqrt{2}$ for atom fields; (2) $\hat a_{c1}=(\hat a_{c} e^{i\varphi_c}+\hat a^\dag_{c} e^{-i\varphi_c})/\sqrt{2}$, $\hat a_{p1}=(\hat a_{p} e^{-i\varphi_p}+\hat a^\dag_{p} e^{i\varphi_p})/\sqrt{2}$, $\hat a_{c2}=(\hat a_{c} e^{i\varphi_c}-\hat a^\dag_c e^{-i\varphi_c})/(\sqrt{2}i)$, $\hat a_{p2}=(\hat a_{p} e^{-i\varphi_p}-\hat a^\dag_{p} e^{i\varphi_p})/(\sqrt{2}i)$ for optical fields; (3) the common and differential modes of incoming optical fields: $\hat a_{\pm\rm in 1/2}=(\hat a_{c\rm in 1/2}\pm\hat a_{p\rm in 1/2})/\sqrt{2}$. Under these basis, the equations of motion Eq.\,\eqref{eq:atomeom} can be transformed to:
\be\label{eq:exacteomApm}
\begin{split}
&(\partial_t\pm i\Omega)\hat A_\pm(t)=\mp\Omega\varphi_{s}\bar A_\mp(t)\\
&\qquad\qquad\qquad\quad\mp i\chi_a[\bar A_\pm(t)\hat a_{+\rm in 1}+i\bar A_{\mp}(t)\hat a_{-\rm in 2}],
\end{split}
\ee
where $\chi_a=\chi\bar a_L$. Then the solution can be written as signal and noise parts, respectively.
The signal part is:
\be
\begin{split}
&A_{s\pm}(t)=\mp\Omega\varphi_{s}\int^t_{t_0} dt'e^{\mp i\Omega(t-t')}\bar A_\mp(t')\\
&\qquad\quad=\mp\varphi_{s}\bar A_\mp(0)\sin{\Omega(t-t_0)}.
\end{split}
\ee
The noise part is:
\be\label{eq:singleaiinout}
\begin{split}
\hat A_{\pm}(t)=e^{\mp i\Omega t}\hat A_\pm(t_0)+\hat A_{\pm\rm opt}(t),\\
\end{split}
\ee
where
\be\label{eq:optnoise}
\hat A_{\pm\rm opt}(t)=\hat A_{\pm \rm am}(t)+\hat A_{\pm \rm ph}(t),
\ee
with
\be\label{eq:opticalnoiseApm}
\begin{split}
&\hat A_{\pm\rm am}(t)=\mp i\chi_a\int^t_{t_0}dt' e^{\mp i\Omega(t-t')}\bar A_\pm(t')\hat a_{+\rm in 1}(t'),\\
&\hat A_{\pm\rm ph}(t)=\pm\chi_a\int^t_{t_0}dt' e^{\mp i\Omega(t-t')}\bar A_\mp(t')\hat a_{-\rm in 2}(t').\\
%&\hat A_{-\rm am}(t)=i\chi_a\int^t_{t_0}dt' e^{i\Omega(t-t')}\bar A_-(t')\hat a_{+\rm in 1}(t'),\\
%&\hat A_{-\rm ph}(t)=-\chi_a\int^t_{t_0}dt' e^{i\Omega(t-t')}\bar A_+(t')\hat a_{-\rm in 2}(t').
\end{split}
\ee
Here, the $e^{\mp i\Omega(t-t')}$ is the free propagator of the atom operators $\hat A_{\pm}$. The $\hat A_{\pm \rm ph}$ and $\hat A_{\pm\rm am}$ are the quantum optical noise contribution to the atom clouds evolution, while the $\hat A_{\pm}(t_0)$ is the initial quantum fluctuation of atom field. 

%This formula can be compared to the Eq.\,\eqref{eq:qmreadout}, except we do not have the back-action term. Interestingly, in Eq.\,\eqref{eq:qmreadout}, such a term is indispensable since the $[\hat y(t),\hat y(t')]=0$ must be satisfied for computer-recorded data string $y(t)$. In Eq.\,\eqref{eq:singleaiinout}, one can easily check that the terms describes the initial quantisation of the probe commute with each other at different times, therefore no back-action term is needed to keep the output field commuting at different times.  

\subsection{Back-action noise}
\begin{figure*}[t]
   \centering
   \includegraphics[scale=0.35]{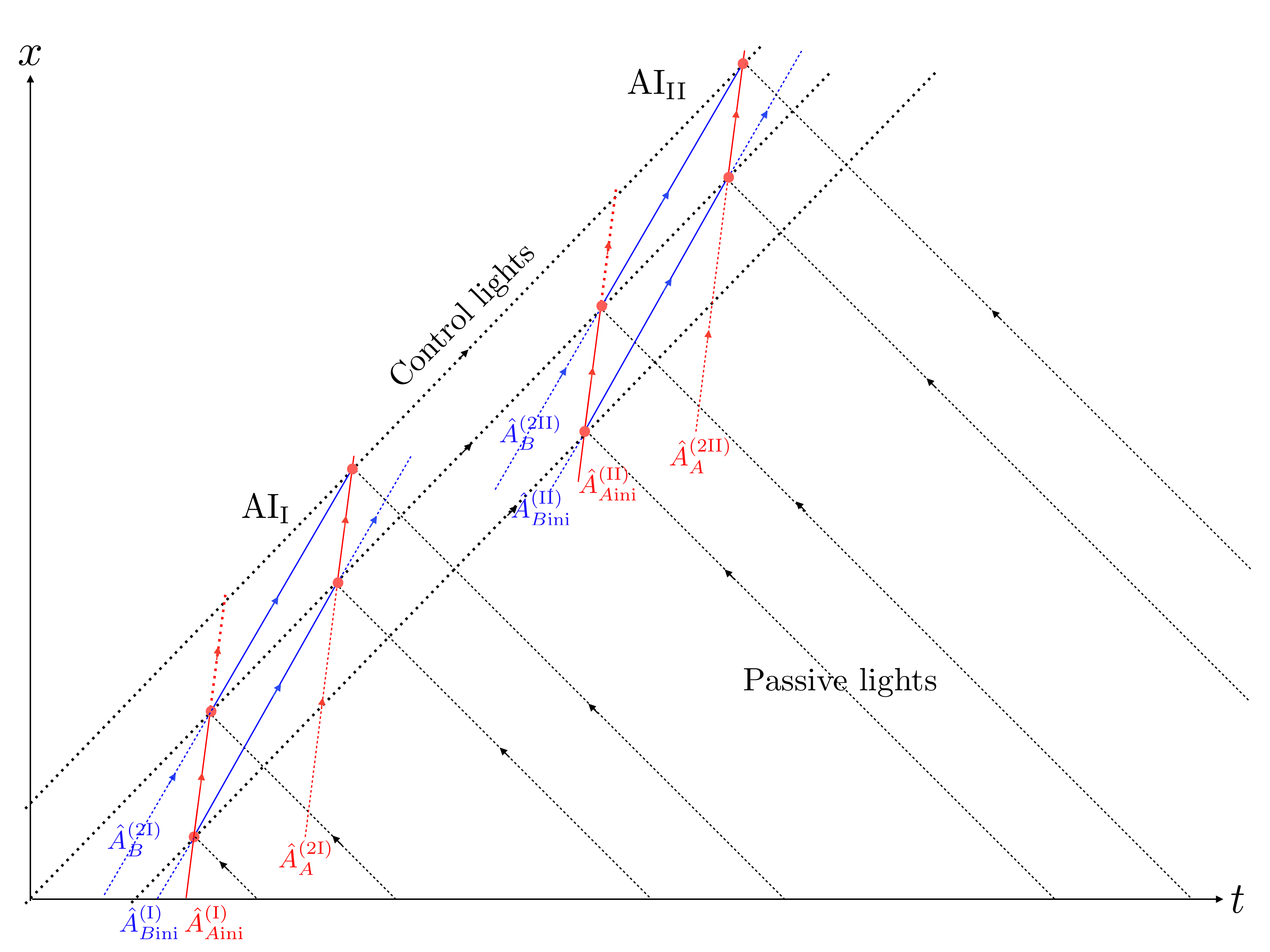} % requires the graphicx package
   \caption{Atom interferometer gravitational wave detector configuration: this is the configuration proposed by Dimopoulos et.al. Two atom interferometers are connected by the control lights. The control fields flying out of the first atom interferometer will carry the corresponding atom informations and impose a back-action noise on the second atom interferometer. The scales in this figure (for illustrative purpose) does not reflects real situation. The distance between two interferometers in the real device is much larger than the length scale of atom interferometers themselves.}
   \label{fig:ai2}
\end{figure*}

For the atom interferometer systems (both for a single atom interferometer and for the GW detector configuration involving a pair of atom interferometers), there exists such situations that the same control fields connects several different interation kernels. For example, in Fig.\,\ref{fig:ai_mz}, the two $\pi/2$ processes are connected by the same control field. In the GW detector configuration proposed by Dimopoulos \emph{et.al} (see Fig.\,\ref{fig:ai2}), all the interaction kernels of the two atom interferometers are connected by the control fields. In these cases, the quantum fluctuation of the first interaction kernel (e.g. denoted by $a$) can be carried by light field (probe) and then affects the second interaction kernel (e.g. denoted by $b$), and finally affects the output atom fields (detector). This would lead to a ``back-action" noise, somewhat similar to the optomechanical system that the quantum fluctuation of light (detector) will be carried by the test masses (probe), and then affect the output light field (detector).  Formally, to analyse such a system, we have to duplicate the Hamiltonian, and the interaction should happen at two different spacetime points:
\be
\begin{split}
H_{\rm int}= \chi \hat A^{(a)\dag}\hat B^{(a)}(\hat a_c^\dag\hat a_d)|^{(a)}+\chi \hat A^{(b)\dag}\hat B^{(b)}(\hat a_c^\dag\hat a_d)|^{(b)}.
\end{split}
\ee

Following the same approach discussed in the last section, one can write down the equations of motion for atom clouds of the second interaction kernel in an almost identical form. The only difference is that the optical fields operators on the r.h.s of the atom equations of motion (Eq.\,\eqref{eq:atomeom}) can be connected to the optical fields flying out of the atom-light interaction region of the first interaction kernel, that is:
\be
\hat a_{c\rm in}^{(b)}=\hat a_{c\rm out}^{(a)}.
\ee

Substituting this relation into the Langevin force Eq.\,\eqref{eq:langevin_force}for the second interaction kernel, and keeping only those terms which due to the atom fluctuations brought from the first interaction kernel, we obtain the ``back-action force'' acting on the atom fields of the second interaction kernel as:
\be
\begin{split}
&[F_{A}^{(b)}]_{\rm BA}=\chi_a^2\bar A_{B}^{(b)}(\bar A_{B}^{(a)*}\hat A_{A}^{(a)}+\bar A_{A}^{(a)}\hat A_{B}^{(a)\dag}),\\
&[F_B^{(b)}]_{\rm BA}=-\chi_a^2\bar A_{A}^{(b)}(\bar A_{B}^{(a)}\hat A_{A}^{(a)\dag}+\bar A_{A}^{(a)*}\hat A_{B}^{(a)}).
\end{split}
\ee
In writing down these expressions, we have used the conditions of balanced control/ passive lasers $|\bar a_p|=|\bar a_c|=a_L$ and the optical field strength for these two interaction kernels are identical $ a_L^{(b)}=a_L^{(a)}=a_L$.

%Substituting this relation into the equation of motion (Eq.\,\eqref{eq:atomeom}) for the second interaction kernel, under the condition of balanced control/ passive lasers $|\bar a_p|=|\bar a_c|=a_L$ and the optical field strength for these two interaction kernels are identical $ a_L^{(b)}=a_L^{(a)}=a_L$, we have:
%\be\label{eq:atomeom2}
%\mathcal{\hat L}
%\begin{bmatrix}
%\hat A_{A}^{(b)}\\
%\hat A_{B}^{(b)}
%\end{bmatrix}
%=-i\chi_a
%\begin{bmatrix}
%\bar A_{B}^{(b)} (\hat a_{c\rm out}^{(a)\dag} e^{-i\varphi_p}+\hat a^{(b)}_{p\rm in} e^{i\varphi_c})\\
%\bar A_{A}^{(b)} (\hat a^{(a)}_{c\rm out} e^{i\varphi_p}+\hat a_{p\rm in}^{(b)\dag} e^{-i\varphi_c})
%\end{bmatrix}.
%\ee
%These equations are written under the condition of balanced passive and control fields:it is easy to prove that the contribution of the first term on the r.h.s of Eq.\,\eqref{eq:opticaleom} vanishes in case $|\bar a_p|=|\bar a_c|=\bar a_L$, and those terms involve $\hat a_{c/p\rm in}$ have exactly the same form as those in Eq.\,\eqref{eq:atomeom}. 
%The terms that we are interested here are those carrying the atom cloud information of the first interaction kernel, namely, the ``back-action terms" given as:
%\be\label{eq:exactbackaction}
%\mathcal{\hat L}
%\begin{bmatrix}
%\hat A_{A}^{(b)}\\
%\hat A_{B}^{(b)}
%\end{bmatrix}_{\rm BA}=(\chi_a)^2
%\begin{bmatrix}
%\bar A_{B}^{(b)}(\bar A_{B}^{(a)*}\hat A_{A}^{(a)}+\bar A_{A}^{(a)}\hat A_{B}^{(a)\dag})\\
%-\bar A_{A}^{(b)}(\bar A_{B}^{(a)}\hat A_{A}^{(a)\dag}+\bar A_{A}^{(a)*}\hat A_{B}^{(a)})
%\end{bmatrix}.
%\ee

Adding these back-action force terms into the atom dynamical equations Eq.\,\eqref{eq:fullatomeom} for the second interaction kernel, intergrating the equations and  expressing these back-action equations in terms of $\hat A_{\pm}, \hat a_{1,2c/p}$, %and integrating the equations, %then under the condition $\bar a_L^{(b)}=\bar a_L^{(a)}=\bar a_L$, 
we have the back-action force contributions $\hat A_{\pm}^{\rm BA}$ as :
\be\label{eq:backaction}
\hat A^{\rm BA}_{\pm}(t)=\mp\frac{\chi_a^2}{2}\int^t_{t_0}dt'e^{\pm i\Omega(t'-t)}[\bar A^{(b)}_{\pm}(t')\mathcal{\hat A}(t')+\bar A_{\mp}^{(b)}(t')\mathcal{\hat B}(t')].
\ee
where
\be
\begin{split}
&\mathcal{\hat A}(t)=\bar A_{+}^{(a)}(t)\hat A_{-}^{(a)\dag}(t)+\bar A_{-}^{(a)*}(t)\hat A_{+}^{(a)}(t)-{\rm h.c},\\
&\mathcal{\hat B}(t)=\bar A_{+}^{(a)}(t)\hat A_{+}^{(a)\dag}(t)-\bar A_{-}^{(a)}(t)\hat A_{-}^{(a)\dag}(t)+{\rm h.c}.
\end{split}
\ee

\section{Interferometry Solution}

This section will give the solution of the atom interferometer. As an example, we only show the solution of which the signal is only contributed from the optical phase imprinted on the atom cloud during the atom-light interaction. This is actually the situation for the proposed atom interferometer GW detectors.  In many other important applications, the signals are carried by  the atom fields themselves. For example, the atom interferometry gravity meter is based on the principle that the gravitational acceleration will affect the propagation phase of the atom fields. In an atom interferometer GW detector, this effect is the physical origin of the gravity noise, which is an important issue that needs to be taken care for the design since the local gravitational field can not be screened. In this paper, we will not discuss these issues (and all the classical noise sources) since they are not the subject of the quantum measurement theory. For simplicity, we also do not consider the effect such as distortion of the atom cloud for simplicity and we assume that the free propagation of atom fields is coherent.

\subsection{Input-output relation}
At the detection stage of an atom interferometer, firstly the particle numbers of A and B atom species are detected respectively, and then the signal is extracted from their difference. Since the detected quantity $\Delta\hat N=\hat N_A-\hat N_B$ is in the $(\hat A_A,\hat A_B)^T$ basis while the formulae of optical noise and back-action terms are more concise in the $(\hat A_+, \hat A_-)^T$ (see Eq.\,\eqref{eq:opticalnoiseApm}), we will use the transformation matrix between these two basis, defined as:
\be
\mathbb{T}(\varphi)=\frac{1}{\sqrt{2}}
\begin{bmatrix}
e^{i\varphi/2}&&e^{i\varphi/2}\\
e^{-i\varphi/2}&&-e^{-i\varphi/2}
\end{bmatrix},
\ee
and the transfer matrix of atom field in the  $(\hat A_A,\hat A_B)^T$ basis is given by:
\be
\mathbb{M}(\theta_j,\varphi_j)=
\begin{bmatrix}
\cos\theta_j&&-i\sin\theta_j e^{i\varphi_j}\\
-i\sin\theta_j e^{-i\varphi_j}&&\cos\theta_j
\end{bmatrix},
\ee
where $\theta_j=\Omega t_j$.\\

For the beam-splitting process (named as step-1), we have:
\be
\begin{bmatrix}
A^{(1)}(t)\\
B^{(1)}(t)
\end{bmatrix}
=
\mathbb{M}(\theta_1,\varphi_{s1}).
\begin{bmatrix}
A(t_0)\\
B(t_0)
\end{bmatrix},
\ee
in which $t_0$ is the initial time of the interrogation process, and $A/B(t)$ can be decomposed into $A/B(t)=\bar A_{A/B}(t)+\hat A_{A/B}(t)$, where $\bar A_{A/B}(t)$ is the mean value of the atom field while $\hat A_{A/B}(t)$ is the perturbation around the mean value. At step-1, $\hat A_{A/B}(t)$ contains the quantum fluctuation of atom field and also the quantum fluctuation of light field, given as:
\be
\begin{bmatrix}
\hat A_{A}^{(1)}(t)\\
\hat A_{B}^{(1)}(t)
\end{bmatrix}
=\mathbb{M}(\theta_1,\varphi_{s1}).
\begin{bmatrix}
\hat A_A(t_0)\\
\hat A_B(t_0)
\end{bmatrix}
+
\mathbb{T}(0).
\begin{bmatrix}
\hat A_{+\rm opt}^{(1)}(t)\\
\hat A_{-\rm opt}^{(1)}(t)
\end{bmatrix}.
\ee

\begin{figure}[h]
   \centering
   \includegraphics[scale=0.23]{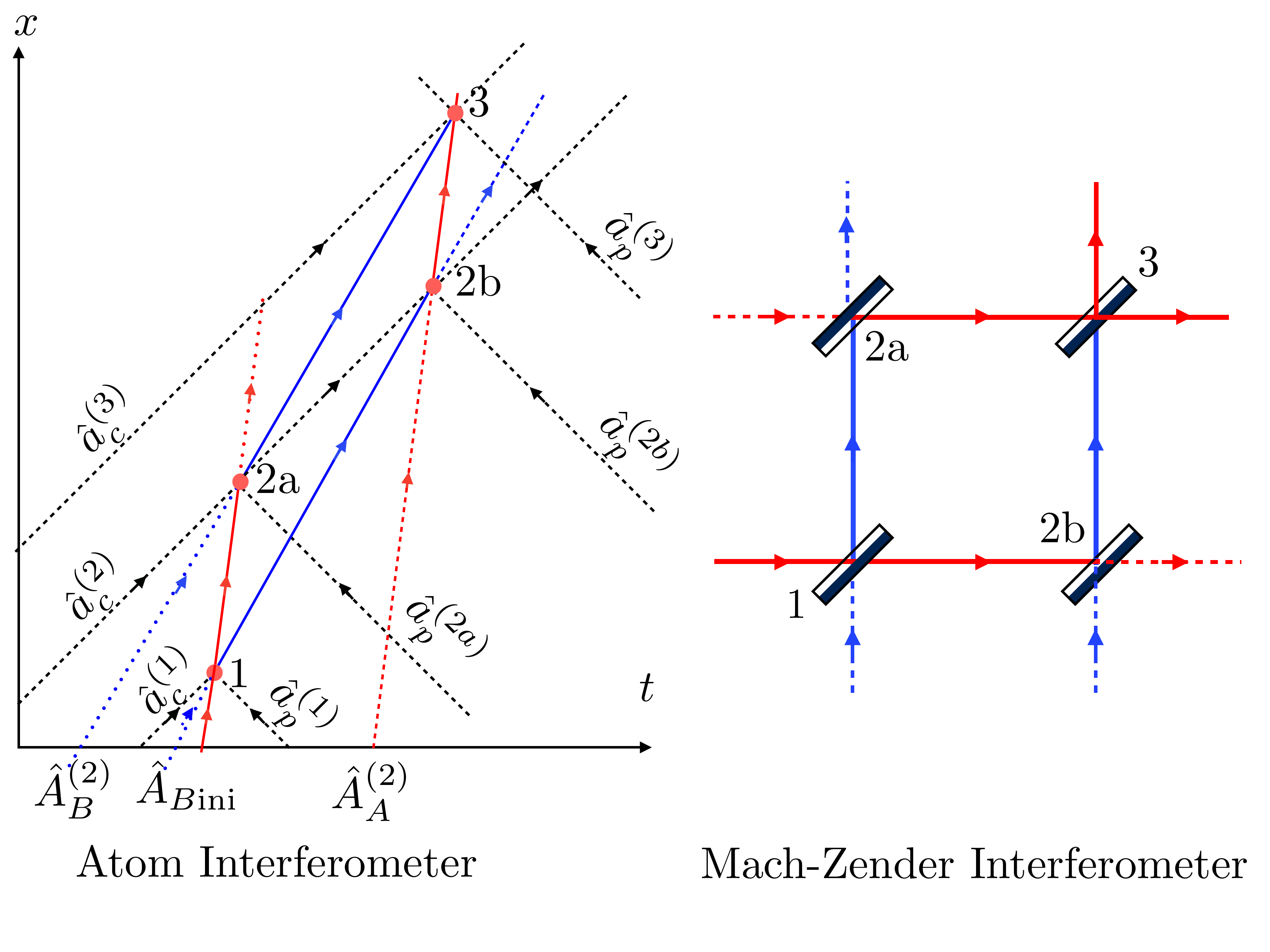} % requires the graphicx package
   \caption{Atom interferometer and optical Mach-Zender interferometer: a comparison. Left pannel: space-time diagram of the $\pi/2$ processes happen in an atom interferometer.Right pannel: an optical Mach-Zender interferometer. On the optical Mach-Zender interferometer, part of the quantum noise injected at 2a and 2b stages are reflected away and left unmeasured. Similar situation also happens in the atom interferometer, where part of the atom noise of $A_{A/B}$ channel injected to 2b/a interaction kernels will be reflected away and left unmeasured. However, the difference is, in the atom interferometer, the ``mirrors" that reflects the matter waves are not uncorrelated as in the optical Mach-Zender interferometer. The  control field that connects the interaction kernel 2a/b is the same field.}
   \label{fig:ai_mz}
\end{figure}

After the step-1, we have $\theta_1=\pi/4$ with $t_1=\pi/(4\Omega)$, and the $A(t)$ and $B(t)$ fields start to separate spatially. The $\pi/2$ processes for $A$-channel and $B$-channel connected by the control light happen sequentially and they should be treated individually. Let us denote the $\pi/2$ processes of the $A$ and $B$ channels to be the step-2a and step-2b, respectively. Clearly, the initial conditions of the step-2a and step-2b processes are $[A^{(1)}(t_1), \hat A^{(2)}_{B}]^T$ and $[\hat A_{A}^{(2)},B^{(1)}(t_1)]^T$ respectively, where $\hat A_{A/B}^{(2)}$ are the field fluctuations injected at the $\pi/2$ steps, shown in Fig.\ref{fig:ai_mz}.

During the $\pi/2$ processes, the corresponding transfer matrices are given by:
\be
\begin{split}
&\text{Step-2a: }\\
&\begin{bmatrix}
A^{(2a)}(t)\\
B^{(2a)}(t)
\end{bmatrix}
=\mathbb{M}(\theta_2,\varphi_{s2}).
\begin{bmatrix}
A^{(1)}(t_1)\\
\hat A^{(2)}_{B}
\end{bmatrix}+
\mathbb{T}(0).
\begin{bmatrix}
\hat A^{(2a)}_{+\rm opt}(t)\\
\hat A^{(2a)}_{-\rm opt}(t)
\end{bmatrix}
,\\
&\text{Step-2b: }\\
&\begin{bmatrix}
A^{(2b)}(t)\\
B^{(2b)}(t)
\end{bmatrix}
=\mathbb{M}(\theta_2,\varphi_{s2}).
\begin{bmatrix}
\hat A^{(2)}_{A}\\
B^{(1)}(t_1)
\end{bmatrix}\\
&\qquad\qquad\qquad\qquad\quad+
\mathbb{T}(0).
\begin{bmatrix}
\hat A^{(2b)}_{+\rm opt}(t)+\hat A^{(2)}_{+\rm BA}(t)\\
\hat A^{(2b)}_{-\rm opt}(t)+\hat A^{(2)}_{-\rm BA}(t)
\end{bmatrix},
\end{split}
\ee
where the upper indices $a/b$ denotes the $A/B$ channels, respectively. Since step-2a and step-2b are connected by the same control light, the control light after step-2a will carry atom information of step-2a and impose a ``back-action" on the step-2b. Here the effect of this back-action is denoted by $\hat A^{(2)}_{\pm \rm BA}(t)$, whose concrete representation can be derived from Eq.\,\eqref{eq:backaction}

%\be
%\begin{split}
%&\hat A^{(2)}_{+\rm BA}(t)=\frac{-\chi_a^2}{2}\int^t_{t_0}dt'e^{i\Omega(t'-t)}[\bar A^b_{+}(t')\mathcal{\hat A}^a(t')+\bar A^b_{-}(t')\mathcal{\hat B}^a(t')],\\
%&\hat A^{(2)}_{-\rm BA}(t)=\frac{\chi_a^2}{2}\int^t_{t_0}dt'e^{i\Omega(t-t')}[\bar A^b_{-}(t')\mathcal{\hat A}^a(t')+\bar A^b_{+}(t')\mathcal{\hat B}^a(t')].\\{}
%\end{split}
%\ee
%where
%\be
%\begin{split}
%&\mathcal{\hat A}^a(t)=\bar A^a_{+}(t)\hat A_{-}^{a\dag}(t)+\bar A_{-}^{a*}(t)\hat A^a_{+}(t)-{\rm h.c},\\
%&\mathcal{\hat B}^a(t)=\bar A^a_{+}(t)\hat A_{+}^{a\dag}(t)-\bar A_{-}^a(t)\hat A_{-}^{a\dag}(t)+{\rm h.c}.
%\end{split}
%\ee\\
%where $\bar A^{a/b}_{\pm}(t)$ are the expectation value of the atom fields of the step-2a/2b in the $A_{\pm}$ basis.

As shown in Fig.\ref{fig:ai_mz}, only one component of the output fields from step-2a/2b participates the recombination stage, while the other component is left unmeasured. For those recombined components, we form a new input field column for the recombination stage as:
$[A^{(2b)}(t_2),B^{(2a)}(t_2)]^T$,where $t_2=\pi/(2\Omega)$.

The fields evolution at the recombination stage now can be written as:
\be\label{eq:inoutatom}
\begin{split}
\left[
\begin{matrix}
A^{(3)}(t)\\
B^{(3)}(t)
\end{matrix}
\right]=
&\mathbb{M}(\theta_3,\frac{\pi}{2}+\varphi_{s3}).
\begin{bmatrix}
A^{(2b)}(t_2)\\
B^{(2a)}(t_2)
\end{bmatrix}\\
&\qquad\quad+
\mathbb{T}(\frac{\pi}{2}).
\begin{bmatrix}
\hat A^{(3)}_{+\rm opt}(t_3)\\
\hat A^{(3)}_{-\rm opt}(t_3)
\end{bmatrix},
\end{split}
\ee
which completes its recombination process at $t=t_3=\pi/(4\Omega)$. Note that this equation can be expanded perturbatively, since the signal terms containing phase $\varphi_s$ and the noise terms are small compared to the expectation values. The results are given as follows.

\noindent$\bullet$ \textbf{Mean field---}Expanding the output atom fields Eq.\,\eqref{eq:inoutatom} to the zeroth order, we obtain the final mean field as:
\be\label{eq:meanfieldinout}
\begin{bmatrix}
\bar A^{(3)}_{A}(t_3)\\
\bar A^{(3)}_{B}(t_3)
\end{bmatrix}=-\frac{1}{\sqrt{2}}
\begin{bmatrix}
e^{i\pi/4}\\
e^{-i\pi/4}
\end{bmatrix}\bar A_A(0).
\ee
\noindent$\bullet$ \textbf{Signal field---} Expanding the output atom fields Eq.\,\eqref{eq:inoutatom} to the first order, we obtain the signal field as:
\be\label{eq:signalfieldinout}
\begin{bmatrix}
A_{As}\\
A_{Bs}
\end{bmatrix}=\frac{1}{2}\bar A_A(0)^2
\begin{bmatrix}
i\varphi_{s1}-(1+i)\varphi_{s2}+\varphi_{s3}\\
-i\varphi_{s1}+(1+i)\varphi_{s2}-\varphi_{s3}
\end{bmatrix}.
\ee
\noindent$\bullet$ \textbf{Atom noise---} Similarly, the noise contributed by the atom fluctuations can be written as:
\be\label{eq:atomfieldinout}
\begin{split}
\begin{bmatrix}
\hat A^{(3)}_{A}(t_3)\\
\hat A^{(3)}_{B}(t_3)
\end{bmatrix}=&\underbrace{-\frac{e^{i\pi/4}}{\sqrt{2}}
\begin{bmatrix}
1&&1\\
-i&&i
\end{bmatrix}
\begin{bmatrix}
\hat A_{A\rm ini}\\
\hat A_{B\rm ini}
\end{bmatrix}}_{\text{atom shot noise}}\\
&\underbrace{-\frac{i\chi_a^2\bar A^2_A(0)}{4\sqrt{2}\Omega}
\begin{bmatrix}
\hat A^{(2)}_{B}+\hat A_{B}^{(2)\dag}\\
-\hat A^{(2)}_{B}-\hat A_{B}^{(2)\dag}
\end{bmatrix}}_{\text{back action noise}}.
\end{split}
\ee
\noindent$\bullet$ \textbf{Optical noise---}The formulae for optical noise are more complicated since they contain contributions from four different steps and the results are:
\be\label{eq:opticalfieldinout}
\begin{split}
&\hat A^{(3)}_{A/B}(t_3)=\frac{1}{2}[-2i\hat A_{+\rm opt}^{(1)}+\hat A_{+\rm opt}^{(2a)}-\hat A_{-\rm opt}^{(2a)}\pm\hat A_{+\rm opt}^{(2b)}+\hat A_{-\rm opt}^{(2b)}\\
&\qquad\qquad\quad+(1\pm i)(\hat A_{+\rm opt}^{(3)}+\hat A_{-\rm opt}^{(3)})],\\
%&\hat A^{(3)}_{B}(t_3)=\frac{1}{2}[-2i\hat A_{+\rm opt}^{(1)}+\hat A_{+\rm opt}^{(2a)}-\hat A_{-\rm opt}^{(2a)}-\hat A_{+\rm opt}^{(2b)}-\hat A_{-\rm opt}^{(2b)}\\
%&\qquad\qquad\quad+(1-i)(\hat A_{+\rm opt}^{(3)}+\hat A_{-\rm opt}^{(3)})].\\
\end{split}
\ee
Substituting the Eqs.\,\eqref{eq:optnoise} and \eqref{eq:opticalnoiseApm} leads to the representation of the above formula in terms of incoming optical noise fields:
\be
\begin{split}
&\hat A^{(3)}_{A/B}(t_3)=\frac{\chi_a\bar A}{2}\int^{\pi/2\Omega}_0dt'\sin{2\Omega t'}[\hat a_{-2}^{(2a)}(t')\pm i\hat a_{-2}^{(2b)}(t')]\\
&\qquad\quad+e^{\pm i3\pi/4}\frac{\chi_a \bar A}{\sqrt{2}}\int^{\pi/4\Omega}_0dt'[\hat a_{+1}^{(3)}(t')+e^{2i\Omega t'}\hat a_{+1}^{(3)}(t')]\\
&\qquad\quad+e^{i3\pi/4}\frac{\chi_a \bar A}{\sqrt{2}}\int^{\pi/4\Omega}_0dt'[\hat a_{+1}^{(1)}(t')+ie^{2i\Omega t'}\hat a_{+1}^{(1)}(t')],\\
%&\qquad\quad+e^{-i3\pi/4}\frac{\chi_a \bar A}{\sqrt{2}}\int^{\pi/4\Omega}_0dt'[\hat a_{+1}^{(3)}(t')+e^{2i\Omega t'}\hat a_{+1}^{(3)}(t')]\\
%&\qquad\quad+e^{i3\pi/4}\frac{\chi_a \bar A}{\sqrt{2}}\int^{\pi/4\Omega}_0dt'[\hat a_{+1}^{(1)}(t')+ie^{2i\Omega t'}\hat a_{+1}^{(1)}(t')].
\end{split}
\ee

\subsection{Standard quantum limit for a single atom interferometer.}
Using Eqs.\,\eqref{eq:meanfieldinout}-\eqref{eq:opticalfieldinout}, we can compute the particle numbers $N_A=A^\dag A$ and $N_B=B^\dag B$ after the recombination completes and expand to the first order of perturbation:
\be
\begin{split}
&N_{A/B}\approx\frac{1}{2}\bar A^2_A(0)\mp\frac{1}{2}\bar A^2_A(0)(\varphi_{s1}-2\varphi_{s2}+\varphi_{s3}),
\end{split}
\ee
Then the $\Delta N=N_B-N_A$, which is the atom number difference at states $|2\rangle$ and $|1\rangle$. , is linearly proportional to the signal:
\be\label{eq:gwsignal}
\Delta N_{\rm signal}=-\bar A^2_{A}(0)(\varphi_{s1}-2\varphi_{s2}+\varphi_{s3}).
\ee

Similar methods can be used to treat the quantum optical noise and the quantum atom noise, the latter of which is given by:
\be\label{eq:atomnoise1st}
\Delta\hat N_{\rm atom}=\underbrace{\bar A_{A}(0)(\hat A_{B\rm ini}+\hat A_{B\rm ini}^\dag)}_{\text{atom shot noise}}+\underbrace{\frac{\chi_a^3\bar A^3_A(0)}{2\sqrt{2}\Omega}(\hat A^{(2)}_{B}+\hat A_{B}^{(2)\dag})}_{\text{back-action noise}}.
\ee
and the optical noise is given by:
\be
\begin{split}
\Delta\hat N_{\rm opt}&=\frac{\bar A_A(0)}{\sqrt{2}}[e^{i3\pi/4}\hat A^{(1)}_{\rm opt-}+e^{i\pi/4}
\hat A^{(1)}_{\rm opt+}-\sqrt{2}\hat A^{(3)}_{\rm opt+}\\
&+\frac{1}{\sqrt{2}}(\hat A^{(2a)}_{\rm opt-}-\hat A^{(2a)}_{\rm opt+}+i\hat A_{\rm opt-}^{(2b)}+i\hat A_{\rm opt+}^{(2b)}]+h.c.
\end{split}
\ee
Now, normalising the particle number difference $\Delta N$ by the signal coefficient,  the estimator of the signal can be written as:
\be
\Delta N_{\rm est}=(\varphi_{s3}-2\varphi_{s2}+\varphi_{s1})+\frac{1}{\bar A^2_A(0)}(\Delta \hat N_{\rm opt}+\Delta\hat N_{\rm atom}),
\ee
Here, we can approximate $\varphi_{s3}-2\varphi_{s2}+\varphi_{s1}\approx \ddot{\varphi}_s T^2$, where $T$ is the interrogation time of the atom interferometer.

To estimate the scaling of the error contributed by these noises, we need to map the parameters in the effective Hamiltonian model to the experimental parameters. It is easy to prove that $\chi_a^2/\Omega=\chi^2|\bar a_L|^2/\Omega=(\Omega a/c)/N_L$ using the relation $\chi |\bar a_L|^2=\Omega$ and the fact that the photon number in the rectangular pulse is $N_L=|\bar a_L|^2l_a/c$ where $l_a$ is the width of the optical pulse. Then the scaling of the error can be estimated as:
\be\label{eq:error}
\begin{split}
\sigma^2_{s}&\sim\frac{N_A}{N_L^2}\left(\frac{\Omega l_a}{c}\right)^2+\frac{2}{N_A}
+\frac{1}{N_L},\\
\end{split}
\ee
in which the first, second and third term are the orders of magnitude of the errors contributed by back-action noise, atom shot noise and purely optical noise, respectively. Apparently the first and second terms have a trade-off when $N_a=N_L$, therefore the error has a minimum value 
\be\label{eq:SQL}
[\sigma^2_{s}]_{\rm mim}\sim \frac{1}{N_L},
\ee
which is actually the photon shot noise (usually $\Omega l_a/c=\pi/2$ or $\pi/4$, i.e. $\Omega l_a/c\sim 1$). This corresponds to the standard quantum limit given in Eq.\,\eqref{eq:sql}. 

It is important to note that this Standard Quantum Limit can only be understood \emph{in the sense of  extrapolation}. Actually when $N_a\sim N_L$, the linear approximation we used in analysing the atom-light interaction will not be valid. The real atom interferometer does not work in this fully-nonlinear region. Therefore for real device, even in the most ideal situation, this Standard Quantum Limit is not accessible as that of LIGO. It only gives a bound to the device sensitivity.

\section{Back-action in atom inteferometer pairs}
The back-action effect discussed in the last section, as we have mentioned, also exists for the system of a pair of atom interferometers. The detector configuration of atom interferometer pair was proposed to measure low frequency GWs. Control fields carrying the atom information of the first interferometer imposes back-action on the second interferometer. The calculation follows the same logic as in the above sections, which is straightfoward but a bit tedious. We only give the final results and discussions here.

The $\Delta \hat N_{\rm atom}$ for the second interferometer is given by:
\be
\begin{split}\label{eq:atomnoise2nd2}
&\Delta \hat N^{(\rm II)}_{\rm atom}=\underbrace{\bar A_A(0)(\hat A^{(\rm II)}_{B\rm ini}+\hat A^{(\rm II)\dag}_{B\rm ini})}_{\text{atom shot noise of ${\rm AI}_{\rm II}$}}+\underbrace{\frac{\chi_a^2\bar A^3_A(0)}{2\sqrt{2}\Omega}(\hat A^{(2\rm II)}_{B}+\hat A^{(2\rm II)\dag}_{B})}_{\text{back action noise inside ${\rm AI}_{\rm II}$}}\\
&\quad\quad\frac{\chi_a^2\bar A^3_A(0)}{2\Omega}[(\hat A^{(\rm I)}_{A\rm ini}+\hat A_{A\rm ini}^{(\rm I)\dag})+i\sqrt{2}(\hat A^{(2\rm I)}_{A}-\hat A_{A}^{(2\rm I)\dag})\\
&\qquad\underbrace{\qquad\quad\quad\quad-(\hat A^{(\rm I)}_{B\rm ini}+\hat A^{(\rm I)\dag}_{B\rm ini})+\sqrt{2}(\hat A^{(2\rm I)}_{B}+\hat A^{(2\rm I)\dag}_{B})]}_{\text{back action noise brought from ${\rm AI}_{\rm I}$}}.
\end{split}
\ee
Here the indices I and II here stand for the first (${\rm AI}_{\rm I}$) and second (${\rm AI}_{\rm II}$) interferometers. The first term in the above $\Delta N_{\rm atom}$ is the atom shot noise of the second interferometer while the second term is the back action noise contributed by atom fluctuations of step-2a of the second interferometer, similar to the second term in Eq.\,\eqref{eq:atomnoise1st}. The last term in Eq.\,\eqref{eq:atomnoise2nd2} represents the back-action imposed by the first interferometer via control light. This result is obtained under the condition that $\bar A^{(\rm II)}_A(0)=\bar A^{(\rm I)}_A(0)=\bar A_A(0)$, that is, the two atom interferometers have the same atom initial states
\footnote{The original expression for the $\Delta N_{\rm atom}$ for the second interferometer is:
$\Delta \hat N^{(\rm II)}_{\rm atom}=
\bar A^{(\rm II)}_A(0)(\hat A^{(\rm II)}_{B\rm ini}+\hat A^{(\rm II)\dag}_{B\rm ini})+(\chi_a^2/2\sqrt{2}\Omega)[\bar A^{(\rm II)}_A(0)]^3(\hat A^{(2\rm II)}_{B}+\hat A^{(2\rm II)\dag}_{B})
+(\chi_a^2/2\Omega)[\bar A^{(\rm II)}_A(0)]^2\bar A^{\rm I}_A(0)[(\hat A^{(\rm I)}_{A\rm ini}+\hat A_{A\rm ini}^{(\rm I)\dag})+i\sqrt{2}(\hat A^{(2\rm I)}_{A}-\hat A_{A}^{(2\rm I)\dag})
-(\hat A^{(\rm I)}_{B\rm ini}+\hat A^{(\rm I)\dag}_{B\rm ini})+\sqrt{2}(\hat A^{(2\rm I)}_{B}+\hat A^{(2\rm I)\dag}_{B})]$, where one can see the ``beating'' of atom fields of these two interferometers. The result shown in the main text is obtained simply by substituting $\bar A^{(\rm II)}_A(0)=\bar A^{(\rm I)}_A(0)=\bar A_A(0)$.}.

The signal field is given by:
\be\label{eq:gwsignal}
\Delta N^{\rm II}_{\rm signal}=\bar A^2_{A}(0)(\varphi^{(\rm II)}_{s1}-2\varphi^{(\rm II)}_{s2}+\varphi^{(\rm II)}_{s3}).
\ee
 
Clearly, the noise described by Eq.\,\eqref{eq:atomnoise2nd2} and Eq.\,\eqref{eq:atomnoise1st} are correlated, since there are terms with $(\hat A^{(\rm I)}_{B\rm ini}+\hat A^{(\rm I)\dag}_{B\rm ini})$ and $(\hat A^{(2\rm I)}_{B}+\hat A^{(2\rm I)\dag}_{B})$ in Eq.\,\eqref{eq:atomnoise2nd2}. This simply means that the two interferometers are \emph{entangled} via the coupling to the same control light fields, if the control light does not decohere strongly during its propagation between two interferometers.

The optical noise in the case of atom interferometer pair consists of the contribution of three control lights and eight passive lights, which is very cubersome and not very interesting in the aspect of quantum measurement theory--- simply contains the initial quantum fluctuation of the probes. We are not going to show it here.

Finally, for extracting the GW signal, we need to substract the measurement results of the two interferometers. Suppose $\varphi_{\rm GW i}= k x_{\rm GW i}$ for a GW-induced optical phase modulation, then $\varphi_{\rm GW1}-2\varphi_{\rm GW2}+\varphi_{\rm GW3}\sim k a_{\rm GW} T^2$, where $a_{\rm GW}=\omega^2h_{\rm GW}L$ is the tidal acceleration (for a monochromatic gravitational wave with frequency $\omega$ and detector baseline length $L$) and $T$ is the interrogation time. A more detailed calculation\,\cite{Dimopoulos2008} showed that the full result (for a monochromatic GW wave with frequency $\omega$ and strain $h_{\rm GW}$) is:
\be
\varphi_{\rm GW}\sim k h_{\rm GW} L \sin^2{\left(\frac{\omega T}{2}\right)}\frac{\sin{\omega L}}{\omega}\sin \omega t.
\ee
in the limit of $\omega T\ll 1$ (which can be easily satisfied form low frequency GW), it reduces to the result here. For the noise part, it is easy to prove that the Eq.\,\eqref{eq:error} and Eq.\eqref{eq:SQL} are still the same in terms of the orders of magnitudes. It is interesting to note that, according to the general theory of linear quantum measurement, there exists a fundamental quantum limit which is the so-called quantum Cramer-Rao bound\,\cite{Miao2017}. Eq.\,\eqref{eq:SQL} is also the  quantum Cramer-Rao bound of the atom interferometer since the signals directly couple to the optical fields (probes) in the TT gauge and the probes' fluctuations here are determined by their own initial quantum states.

\section{Dynamics of the effective operators --- a more exact treatment}
The exact definition of the operators $\hat A_A,\hat A_B$ etc, and the coupling strength $\chi$ in the effective Hamiltonian can be determined by using a field theory approach developed in the Appendix. This field theory approach is based on the following action:
\be
S_{\rm int}=g\int d^2x \phi_{A}(x)\phi_{B}(x)\phi_{c}(x)\phi_{p}(x),
\ee
where the coordinates $x$ represents $(t,z)$. The relationship between $g$ and the physical quantities describing the atom-light interaction such as the atom dipole moments, frequencies of different energy levels, etc. is given in the appendix.  The corresponding equations of motion are given by:
\be
\begin{split}
&(\partial_t+v_{A}\partial_z)\tilde\phi^{+}_{A}=g_A\tilde\phi^+_{B}\tilde\phi^+_p\tilde\phi^-_c,\\
&(\partial_t+v_{B}\partial_z)\tilde\phi^{+}_{B}=g_B\tilde\phi^+_{A}\tilde\phi^-_p\tilde\phi^+_c,\\
&(\partial_t+\partial_z)\tilde\phi^+_{c}=g_c\tilde\phi_A^-\tilde\phi_B^+\tilde\phi^+_{p},\\
&(\partial_t-\partial_z)\tilde\phi^+_{p}=g_p\tilde\phi_A^+\tilde\phi_B^-\tilde\phi^+_{c}.
\end{split}
\ee
Here the $\tilde{\phi}^{+}_j$ ($j=A,B,c,p$) are the slowly varying amplitude field operators of the positive branch defined through $\hat{\phi}_{j}^+(x)=\tilde{\phi}^+_{j}(x)e^{-i \omega_{j0}(t-t_{j0})+ik_{j0}(z-z_{j0})}$ where $\omega_{j0}$ are the frequency of the free fields $\hat\phi_i$ and related to the wave vector $k_{j0}$ through $\omega_{j0}^2=k_{j0}^2+m_{j0}^2$ (for optical fields, the masses are zero). The $v_{A/B}$ is the WKB velocity of atom wave packet A/B and the two optical fields are propagating along the opposite directions. The coupling constants are defined as: $g_j=ig/(2\omega_{j0})$. The $\hat{\phi}_j^-$ is the corresponding negative branch field operators.   As shown in details in Appendix, the initial states of the mean optical fields can be treated as plane waves, the initial states of the mean atom fields are zero and a Gaussian profile for the $\phi_B$ and $\phi_A$, respectively. This Gaussian profile, in the spacetime-domain is given by:
\be
\bar\phi_A=\frac{\bar \alpha_A\Delta_A^{1/2}}{(2\pi)^{1/4}}{\rm exp}\left[-\frac{\Delta_A^2}{4}[z-z_0-v_a(t-t_0)]^2\right],
\ee
where the $\bar\alpha_A$ is the coherent amplitude and the $1/\Delta_A$ is the width of the Gaussian profile.

\subsection{Perturbative solution to the optical fields: effective operator for atoms}
Typically, in an interferometric process, the light-atom interaction time is very short compared to the free evolution time of the atom cloud, and the centre of mass velocity of the atom cloud is very low, typically $\sim 2$\,cm/s. Therefore to the leading order, we can treat the atom center of mass motion to be static during the interaction process, that is, $v_A\approx v_B\approx 0$. We also note that the spatial size of optical fields are much larger than the size of the atom cloud, therefore we can approximate the mean value of the optical fields to be almost constants during the interaction process. For this calculation, we only care about control field because it transfers noise to the next atom-light interaction kernel, while different kernels interact with different passive fields, as shown in Fig.\,\ref{fig:ai_mz}.

\begin{figure}[h]
   \centering
   \includegraphics[scale=0.25]{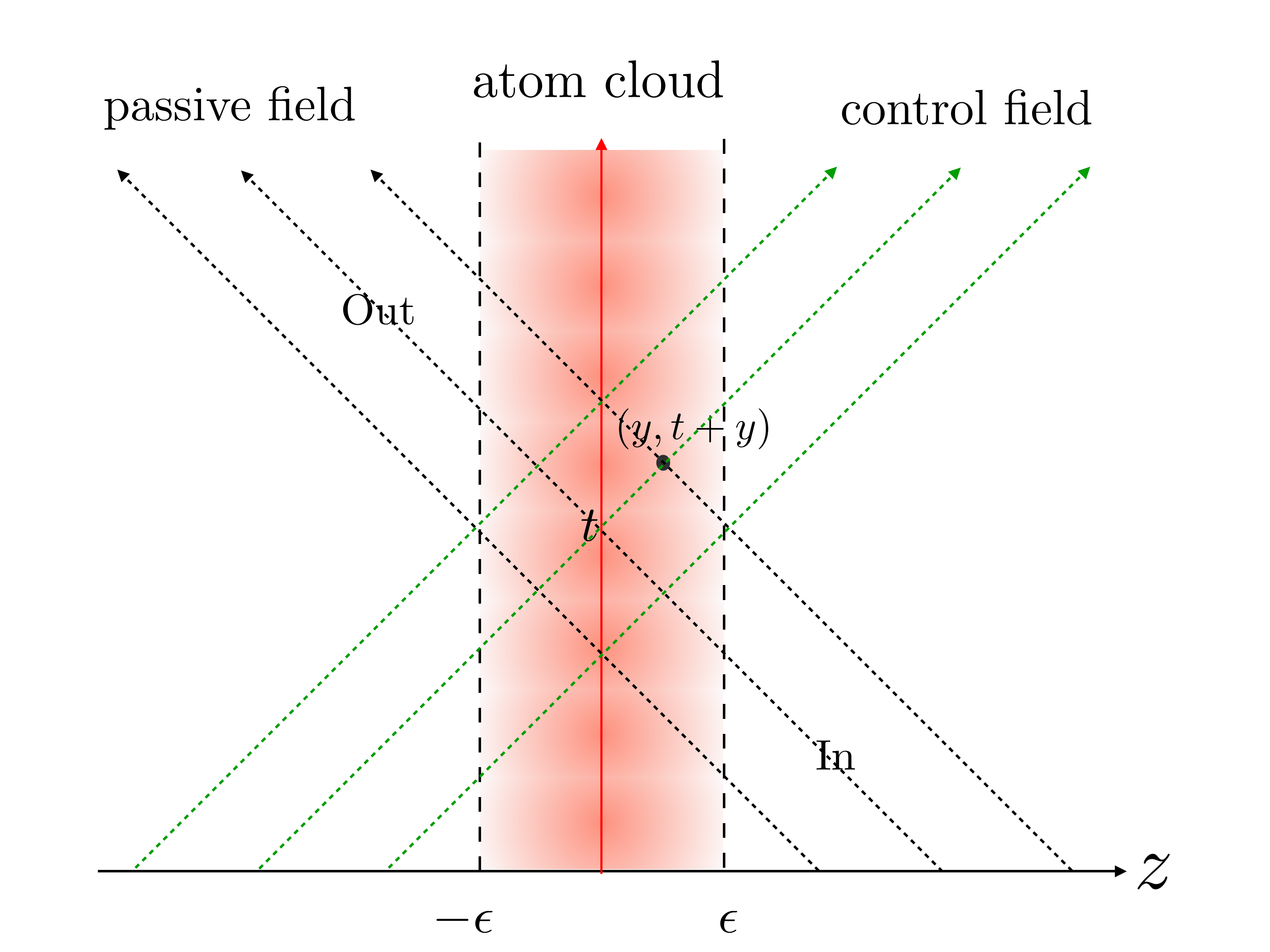} % requires the graphicx package
   \caption{Atom-light interaction kernel. The atoms are undergoing the state swapping interaction, and during the very short interaction time (typically $\sim \mu$s), the A and B atoms does not have enough time to fly apart. Our effective atom operators $\Phi$ are defined through integration over the atomic cloud profile on the spatial direction. The discussion in the Section II will reduce the problem to be an effective model describing the interaction of the effective atom operators with the incoming light fields.}
   \label{fig:kernel}
\end{figure}

These equations can be solved in a perturbative way. For the equation of motion of the control field, the first order perturbation formal solution is given by:
\be
\begin{split}
\delta &\phi_c^+(t+z,z)-\delta\phi_c^+(t-\epsilon,-\epsilon)=\\
&g_c\int^z_{-\epsilon}dy\delta[\phi_A^-(t+y,y)\phi_B^+(t+y,y)\phi_p^+(t+y,y)],
\end{split}
\ee
where the atom cloud mostly distributed in $[-\epsilon,\epsilon]$, as shown in Fig.\,\ref{fig:kernel} (we move the coordinate origin to the atom center of mass position).
For brevity, we remove the tilde on the operators. In the following, all operators are the slowly varying amplitude operators. Expanding the right hand side to the first order, we obtain:
\be
\begin{split}
&g_c\int^z_{-\epsilon}dy\bar \phi_p^+\bar \phi_B^+(t+y,y)\delta \phi_A^-(t+y,y)\\
&+g_c\int^z_{-\epsilon}dy\bar \phi_p^+\bar \phi_A^-(t+y,y)\delta \phi_B^+(t+y.y)\\
&+g_c\int^z_{-\epsilon}dy\bar \phi_B^+(t+y,y)\bar \phi_A^-(t+y,y)\delta \phi_p^+(t+y,y).
\end{split}
\ee
The classical atomic fields can be written as (under the slow motion approximation):
\be
\bar \phi^{+(-)}_{A/B}(t,y)=f_a(y)\bar \alpha^{(*)}_{A/B}(t),
\ee
where 
\be
f_a(y)=\frac{\Delta_a^{1/2}}{(2\pi)^{1/4}}{\rm exp}\left[-\frac{1}{4}\Delta_a^2y^2\right].
\ee

Since $y\in[-\epsilon,\epsilon]$ and $\epsilon\ll1$, we can expand 
\be
\bar \alpha_{A/B}(t+y)\approx\bar \alpha_{A/B}(t)+y\dot{\bar \alpha}_{A/B}(t).
\ee
Note that $|y\dot{\bar \alpha}_{A/B}(t)|\sim\Omega y \bar \alpha_{B/A}(t)$ and $\Omega y\ll 1$, we can simplify the above terms to be:
\be
\begin{split}
&g_c\bar\phi_p^+\left[\bar \alpha_B(t)\int^z_{-\epsilon}dyf_a(y)\delta \phi_A^-(t,y)+\bar \alpha_A^*(t)\int^z_{-\epsilon}f_a(y)\delta \phi_B^+(t,y)\right]\\
&+g_c\int^z_{-\epsilon}dy\bar \alpha^*_A(t)\bar \alpha_B(t)f_a^2(y)\delta\phi_p^+(t+y,y).
\end{split}
\ee

Now let us define the \emph{effective atom operators} to be:
\be\label{eq:atomoperator}
\delta\Phi^{\pm}_{A/B}(t)={\rm lim}_{z\rightarrow\epsilon}\int^z_{-\epsilon}dyf_a(y)\delta\phi^{\pm}_{A/B}(t,y).
\ee
As we shall see later, these effective operators have a nice property that \emph{the commutation relation of the associated creation and annihilation operators normalises to one.} The Physical interpretation of the effective operators is that it describes the whole wavepacket of the atom field. Using these effective operators, the input-output relation for a light ray passing through the atomic cloud can be written as (for the passive field, it can be calculated in the same way):
\be\label{eq:opticalinoutphi}
\begin{split}
&\delta \phi_{c\rm out}^+(t)-\delta\phi_{c\rm in}^+(t)=g_c\bar \phi_p^+[\bar \alpha_B(t)\delta\Phi^-_{A}(t)+\bar \alpha^*_A(t)\Phi^+_B(t)]\\
&\qquad\qquad+g_c\bar \alpha_A^*(t)\bar \alpha_B(t)\int^\epsilon_{-\epsilon}dyf_a^2(y)\delta\phi_p^+(t+y,y),\\
&\delta \phi_{p\rm out}^+(t)-\delta\phi_{p\rm in}^+(t)=g_p\bar \phi_c^+[\bar \alpha^*_B(t)\delta\Phi^+_{A}(t)+\bar \alpha_A(t)\Phi^-_B(t)]\\
&\qquad\qquad +g_p\bar \alpha_A(t)\bar \alpha_B^*(t)\int^\epsilon_{-\epsilon}dyf_a^2(y)\delta\phi_p^+(t-y,y).
\end{split}
\ee
The negative frequency branches simply obey the Hermitian conjugate of the above equations. The ratio between the second term and the first term on the r.h.s of the equation is $\sim\sqrt{N_a/N_L}\ll1$, therefore can be safely ignored.

Furthermore, we introduce the creation and annihilation operators that correspond to those effective atom operators and also the optical operators.  Remind that the operators here are related to the creation and annihilation operators as follows:
\be
\delta\Phi_{A/B}^+=\frac{\hat A_{A/B}(t)}{\sqrt{2\omega_{a}}},\quad \delta\phi_{c/p\rm in}^{+}=\frac{\hat a_{c/p\rm in}(z, t)}{\sqrt{2\omega_{L}}},
\ee
where we take assumptions $\omega_{c0}\approx\omega_{p0}=\omega_L$ and note that $\bar \Phi^+_{A/B}=\bar \alpha_{A/B}/(2\omega_a)$. Now we want to rewrite the equations of motion of the atom fields in a more concise way, using these creation and annihilation operators. First, let us check the dimension of the above defined creation and annihilation operators. We have 
\be
\begin{split}
&\hat A_{A/B}(t)=\int dz f_a(z)\hat a_{A/B}(z,t),\quad \text{and}\\
&[\hat a_{A/B}(z,t),\hat a^\dag_{A/B}(z',t)]=\delta(z-z').
\end{split}
\ee
where the commutation relation here is the standard one on a time slice $t$. Using this commutation relation and the normalisation condition for $f_a(z)$, it is straightforward to show that $[\hat A_{A/B}(t),\hat A^\dag_{A/B}(t)]=1$, and $\hat A$ is a dimensionless operator, $\hat a_{c\rm in}(z,t)$ has the dimension $[{\rm length}]^{-1/2}$.  The gravitational wave community is more familiar with the operator satisfying $[\tilde a(t), \tilde a^\dag(t')]=\delta (t-t')$, it is important to note that this is an equal time commutation relation for propagating fields, and the operators here are related by $\tilde a_{c\rm in}/\hat a_{c\rm in}=c^{1/2}$ where $c$ is the speed of light. In the following, we will use the $\tilde a$ and replace tilde with hat. Then the Eq.\,\eqref{eq:opticalinoutphi} can be translated to:
\be\label{eq:opticalinouta}
\begin{split}
&\hat a_{c\rm out}-\hat a_{c\rm i}=i\chi_L e^{-i\varphi_p}[\bar \alpha_B(t)\hat A_A^\dag+\bar \alpha_A^*(t)\hat A_B],\\
&\hat a_{p\rm out}-\hat a_{p\rm i}=i\chi_L e^{-i\varphi_c}[\bar \alpha^*_B(t)\hat A_A+\bar \alpha_A(t)\hat A_B^\dag].
\end{split}
\ee
where we have $\chi_L=|g_c|\bar\phi_L\sqrt{\omega_L/\omega_a}$ under the approximation that $\omega_c=\omega_p=\omega_L$ and $|\bar\phi_p^+|=|\bar\phi_c^+|=\bar\phi_L=\bar a_L/\sqrt{2\omega_L}$. Comparing this equation with Eq.\,\eqref{eq:opticaleom}, we know that the $\chi$ in the effective Hamiltonian has the form:
\be\label{eq:mapping}
\chi\rightarrow-g/(2\omega_L)=g_c=g_p.
\ee

\subsection{Deriving the evolution of atom fields using field theory}
Using Eq.\,\eqref{eq:atomoperator}, we integrate the perturbation equations of atom fields to obtain the perturbation equations of \emph{effective atomic operators}:
\be
\begin{split}
&\partial_t\delta \Phi_A^++i\Omega e^{i\varphi}\delta\Phi_B^+=g_a\bar\phi^+_p\int^\epsilon_{-\epsilon}dzf_a(z)\bar\Phi^+_B(t,z)\delta\phi_c^-(t,z)\\
&\qquad\qquad\qquad\quad+g_a\bar\phi^-_c\int^\epsilon_{-\epsilon}f_a(z)\bar\Phi^+_B(t,z)\delta\phi_p^+(t,z),\\
&\partial_t\delta \Phi_B^++i\Omega e^{-i\varphi}\delta\Phi_A^+=g_a\bar\phi^-_p\int^\epsilon_{-\epsilon}dzf_a(z)\bar\Phi^+_A(t,z)\delta\phi_c^+(t,z)\\
&\qquad\qquad\qquad\quad+g_a\bar\phi^+_c\int^\epsilon_{-\epsilon}f_a(z)\bar\Phi^+_A(t,z)\delta\phi_p^-(t,z).\\
\end{split}
\ee
Now, we substitute the formal solution of $\delta \phi^{\pm}_{c/p}$ into the above equations, which will reveal the structure of the optical back-action on the atom fields.

Let us take the first term on the r.h.s of $\delta \Phi_A$ equation as an example, it has the following form after substituting $\delta\phi_c^-(t,z)$:
\be
\begin{split}
&g_a\bar\phi^+_p\int^\epsilon_{-\epsilon}dzf_a(z)\bar\Phi^+_B(t,z)\delta\phi_c^-(t,z)=g_a\bar\phi^+_p\alpha_B(t)\times\\
&\int^\epsilon_{-\epsilon}dzf^2_a(z)\left\{\delta\phi_{c\rm in}^-(t)+g_c\alpha_A(t)\alpha_B^*(t)\int^z_{-\epsilon}dyf_a^2(y)\delta\phi_p^-(y)+\right.\\
&\left.g_c\bar\phi_p^-\int^z_{-\epsilon}dyf_a(y)[\alpha_A(t)\delta\phi_B^{-}(y,t)+\alpha_B^*(t)\delta\phi_A^+(y,t)]\right\}.
\end{split}
\ee
Note that the $f_a(y)$ takes a Gaussian form symmetric around $y=0$, which means that we can write:
\be
\int^\epsilon_{-\epsilon}dz f_a^2(z)\int^z_{-\epsilon}dy f_a(y)\delta \phi_A^+(y,t)=\frac{1}{2}\delta\Phi_A^+(t),
\ee
where we have used the normalisation condition for $f_a(z)$, and we have:
\be
\begin{split}
&g_a\bar\phi^+_p\int^\epsilon_{-\epsilon}dzf_a(z)\bar\Phi^+_B(t,z)\delta\phi_c^-(t,z)=g_a\bar\phi^+_p\alpha_B(t)
\delta\phi_{c\rm in}^-(t)\\
&\qquad\frac{1}{2}|g_ag_c||\bar\phi_p|^2[\alpha_A(t)\alpha_B(t)\delta\Phi_B^{-}(t)+|\alpha_B(t)|^2\delta\Phi_A^+(t)].
\end{split}
\ee
Here, we ignore the $\bar \phi_p^-$ term since it is much smaller than the other terms.
In a similar way, we have:
\be
\begin{split}
&g_a\bar\phi^-_c\int^\epsilon_{-\epsilon}dzf_a(z)\bar\Phi^+_B(t,z)\delta\phi_p^+(t,z)=g_a\bar\phi^-_c\alpha_B(t)
\delta\phi_{p\rm in}^+(t)\\
&\qquad-\frac{1}{2}|g_ag_c||\bar\phi_c|^2[\alpha_A(t)\alpha_B(t)\delta\Phi_B^{-}(t)+|\alpha_B(t)|^2\delta\Phi_A^+(t)].
\end{split}
\ee

It is clear that those terms containing $\delta \phi_{c/p{\rm in}}$ are the optical noise while the rest containing $\delta \Phi_{A/B}$ is the dynamical back-action of the control/passive optical fields to the atom dynamics, which is a manifestation of fluctuation-dissipation theorem. Note that when $|\bar \phi_c|=|\bar \phi_p|=|\bar \phi_L|$, the dynamical back-action contributed by passive and control beam would cancel with each other, leaving only the optical noise.

Finally, we have the equations of motion for atom fields as:
\be
\begin{split}
&\partial_t\delta \Phi_A^++i\Omega e^{i\varphi}\delta\Phi_B^+=\\
&\qquad\qquad\qquad g_a\bar\phi_L\alpha_B(t)[e^{-i\varphi_p}\delta\phi_{c\rm in}^-(t)+e^{i\varphi_c}\delta\phi_{p\rm in}^+(t)],\\
&\partial_t\delta \Phi_B^++i\Omega e^{-i\varphi}\delta\Phi_A^+=\\
&\qquad\qquad\qquad g_a\bar\phi_L\alpha_A(t)[e^{i\varphi_p}\delta\phi_{c\rm in}^+(t)+e^{-i\varphi_c}\delta\phi_{p\rm in}^-(t)].
\end{split}
\ee

Then we can obtain the exact form of equations of motion for atom cloud:
\be\label{eq:exacteomAB}
\begin{split}
&\partial_t\hat A_A+i\Omega e^{i\varphi} \hat A_B=i\chi_B(t)[e^{-i\varphi_p}\hat a_{c\rm in}^\dag(t)+e^{i\varphi_c}\hat a_{p\rm in}(t)],\\
&\partial_t\hat A_B+i\Omega e^{-i\varphi}\hat A_A=i\chi_A(t)[e^{i\varphi_p}\hat a_{c\rm in}(t)+e^{-i\varphi_c}\hat a_{p\rm in}^\dag(t)].\\
\end{split}
\ee
where $\varphi=\varphi_c-\varphi_p$ is the phase difference between control and passive lights.
\be
\chi_{A/B}(t)=\sqrt{\frac{\omega_a}{c\omega_L}}|g_a|\bar\phi_L\alpha_{A/B}(t).
\ee
The $\alpha_{\pm}(t)=\alpha_{\pm}(0){\rm exp}[\mp\Omega t]$ and $\chi_0:=\sqrt{\omega_a/c\omega_L}|g_a|\bar \phi_L$. Using the relation Eq.\,\eqref{eq:mapping}, we can also map Eq.\,\eqref{eq:exacteomAB} to Eq.\,\eqref{eq:atomeom}.

\section{Comparison with the Laser interferometer GW detector}
Now we can make some comparison between Laser Interferometer GW detector and atom interferometer detector from several different aspects:

\textbf{$\bullet$ Discrete or Continuous---}LIGO is a detector where the optical field continuously monitors the position of the test masses, recording the continuous waveform of the gravitational wave. The continuity of such measurement is the reason for the existence of SQL. However, the atom interferometer works differently, in a somewhat discretised way. Basically, each interrogation process records one data point of the waveform time-series, and the measurements of different data points are mutually independent. To record the waveform of the GWs, the interrogation process needs to be repeated many times. For each data point, the measurement is continuous, with the time scale equal to the interrogation time scale of the matter wave interferometry. Therefore, the quantum limit discussed here is the limitations to the data point recorded by each individual interrogation process.

\textbf{$\bullet$ Measurement quantity---} Gravitational wave information is carried in the curvature perturbation $\Psi\sim \ddot{h}$, which corresponds to the acceleration of the test masses. For LIGO, we have the equation of motion for the test masses as $m\ddot{x}\sim m\Psi\sim mL\ddot{h}$, where $m$ and $L$ are effective mass of the test masses motion and the baseline length of the interferometer, respectively. The light field will directly carry the information of test mass displacement and the interferometer is a displacement sensor. However, for an atom interferometer, each interrogation process directly records the acceleration as shown in Eq.\eqref{eq:gwsignal}, therefore, an atom interferometer is an acceleration sensor.

\textbf{$\bullet$ Test mass quantisation---}As we have shown in Eq.\,\eqref{eq:qmreadout}, test mass quantisation will generally have an impact on the measurement result. LIGO's measurement result will be in principle affected by the test mass quantisation if we directly apply Eq.\,\eqref{eq:qmreadout}. However, if we do the post-data processing to extract the curvature perturbation (acceleration) information, we are targeted on $x(2\Delta t)-2x(\Delta t)+x(0)$. Using the free mass evolution equation $\hat x(t)=Lh(t)+\hat p_0t/m+\hat x_0$, it is easy to prove that all the information about the initial test mass quantum state will be eliminated when we extract the acceleration information. This important result has been obtained by Braginsky {\it et al.}\,\cite{Braginsky2003} The key for the elimination of test mass quantisation effect is the fact that the differential motion of th test masses of the two arms is a single degree of freedom during the entire detection process. However, for atom interferometer (Dimopolous configuration), four different pairs of laser beams are needed to complete one interrogation period and they belong to different degrees of freedom. Therefore, the probe quantisation effect can not be removed in the same way as LIGO, that is why we need to consider the quantum fluctuation of light field (or the probe in a more general sense) in discussing the sensitivity of the atom interferometer.

\textbf{$\bullet$ Back-action---}In LIGO, the back-action is contributed by the radiation pressure force acting on the test mass. In the atom interferometer, the back action comes from the atom noise carried by the control fields linking two atom interferometers. In LIGO, the back action can not only be a noise source, but also change the dynamical behavior of the system in certain parameter regions. For example. If the resonant frequency of the interferometer does not match the carrier laser frequency, which can be done through tuning the signal recycling cavity, the dynamics of the test mass and optical field will change and create an optical rigidity for the test mass. In atom interferometer, the state transition dynamics (Rabi rate) of the atom cloud can also be changed by the interaction between optical and atomic fields, if there is an intensity unbalance between the control and passive fields, as we have shown in details in Section II.A and Section V.B.

%\textbf{$\bullet$ Fundamental Quantum limit---}In LIGO, we have the fundamental quantum limit (FQL) (or Cramer-Rao bound)\,\cite{Miao2017} with noise spectrum $S^{\rm SQL}_{xx}(\Omega)=\hbar^2c^2/2S_{PP}(\Omega)$ where $S_{PP}(\Omega)$ are the spectrum of power fluctuation of intra-cavity fields. Apparenly, these two limits have different forms. However, we find that the SQL and FQL in the atom interferometer takes the same form, i.e., they all $\sim 1/N_L$. The reason behind this difference can be understood as follows. In LIGO, GWs couples to the probe (test masses) in the local Lorentz gauge and quantum Cramer-Rao bound tells us that the FQL should be determined by the initial fluctuation of the test masses. However, the probe (test masses) are coupled to the detector (optical field) and the initial quantisation of the test mass is determined by the optical fluctuations. Therefore, the FQL is determined by $S_{PP}(\Omega)$. In the atom interferometer, the coupling between detector (atom) and probe (light) is so weak that the probe's initial fluctuation is still governing the 

The above discussion qualitatively compares the physics of atom interferometer GW detector with the LIGO detector. It also worthes to comment the similarity between atom interferometer GW detector with the LISA detector. Unlike the Michelson type LIGO detector, the LISA detector is actually a transponder system where pairs of test masses are connected by an optical link and the optical phase change due to the gravitational waves are recorded with local interferometry set up around each test mass. Atom interferometer also matches with this picture, where two atom clouds are connected by an optical link and the phase change of this optical link is recorded. In LISA, only the optical sensing noise is worth to be considered and the quantum back action is extremely weak since the optical power recieved on each satellite is of picowatts. Zero-point fluctuation of the test masses on LISA  can be eliminated since it is also a displacement sensor. For a detailed study for the comparison between atom interferometer GW detector with LISA detector, see\,\cite{Baker2012,Bender2012}

\section{Discussion and Conclusion}
In this paper, the matter wave interferometry, in particular the example of a 
one-dimensional model of atom interferometer, is carefully added onto the jigsaw puzzle of the linear quantum measurement theory. Previous studies on the atom interferometry\,\cite{Borde2001,Dimopoulos2008prd}, although quite complete and careful, mostly worked in the Schr{\"o}dinger picture and did not speak in the langauge of quantum measurement theory. Establishing such a theory for matter wave interferometry can help the LIGO community clearly understand the physics of atom interferometer GW detectors. Our result demonstrates in detail about how the light-atom interaction affects the dynamical and noise behavior of atom cloud and gives the input-output relation for the linear response of the device. Concretely, we clarify how the concepts of detector shot noise, probe's zero-point fluctuation, back-action noise and dynamical back-action manifest themselves in the atom interferometry. Similar to LIGO case, we also obtain the formula for the Standard Quantum Limit in the atom interferometry.

The configuration raised by Dimipolous \emph{et.al} is the most original one proposed for detecting GWs. Other configurations were also raised\,\cite{Yu2011,Graham2013,Canuel2018}, in particular, the so called single-photon atom interferometer, where the interferometric process happen via transition of Rabi oscillation of two-level atoms rather than Raman transition\,\cite{Yu2011}. The advantage of this configuration is that the optical noise (zero-point fluctuations of probe) can be removed in the ideal one dimensional case through common mode rejection. In order to increase the fringe visiblity of the atom interferometer, the distinguishbiliy of the atom cloud trajectory should be increased which means we need a larger momentum transfer\,\cite{Mueller2008lmt,Clade2009,Onur2016}. The formalism developed in this paper can be extended to these configurations also. Moreover, the concept of optical-cavity assisted atom interferometry was also discussed and experimentally tested\,\cite{Hamilton2015cavity, Canuel2018}, using our formalism to study these configurations will be a future work.

The analysis in this work is targeted on revealing the physical principle. Therefore we focus on the simplest case where we ignore the classical noise sources and optical losses that would more seriously affect the sensitivity of atom interferometry. One typical example is, the back-action noise carried by the control fields that propagate from one interferometer to the other will not contribute to the sensitivity in the current design of atom interferometer GW detector. The reason are the following. (1) the real three-dimensional atom-light interactions happen in a point-scattering way that the cross sectional area of the light field is much larger than the that of the atom field. Therefore such a scattering itself is very lossy thereby noisy. (2) for detecting GWs, these two interferometers must be separated in a relatively large distance so that the atom cloud of the second interferometer will only interacts with a small patch of the large wavefront propagated from the first interferometer. The diffraction loss will erase most of the information of the atom cloud of the first interferometer carried by the light field. Here, we want to emphasize that the lose of back-action noise here does not imply that the current atom-interferometer designs are perfect, quite contrary, it implies that current designs are too lossy to have the back-action issue.  It is probably insightful to study the method to mitigate these issues in the future. General understanding tells us that the capability of any type of interferometric experimental platform is determined by the coherence of every physical steps in the device and the goal for reaching a better sensitivity is practically realised by mitigating all the issues that decohere the waves in the interferometer. To mitigate the first issue mentioned above, we have to design the system so that we have a mode-matched atom-light interaction and thereby a more coherent scattering of light by atom cloud, which is very difficult under the current technology; while to mitigate the second issue, one way is to place the two atom interferometers in some optical cavity assisted structures. This idea has been proposed for the MIGA (Matter-wave laser Interferometric Gravitation Antenna) project\,\cite{Canuel2018}.  Analysing the quantum noise for such a device will be a future extension of this work.

\section*{acknowledgement}
Y.M. thanks Zebin Zhou for his constant support and encouragement, especially for sharing his interests in this work during the trip to Hanford. Y.M. also thanks Albert Roura for a fruitful discussion on different types of atom interferometer and Haixing Miao for an insightful discussion on the fundamental quantum limit. Y.M. thanks Yuta Michimura and Kentaro Somiya for discussions on the concepts of using atom interferomter to detect gravitational waves. Y.C. thank Rana Adikhari for a discussion on the experimental feasibility. He also thanks Mark Kasevich and Hoger M\"{u}ller for interesting discussions on this work. Y.M, X.L, and Y.C. thank Belinda Pang for discussions on the early stage of this paper. Y.C., X.L., and Y.M. are supported by the National Science Foundation through Grants PHY- 1612816, PHY-1708212 and PHY-1708213, the Brinson Foundation, and the Simons Foundation (Award Number 568762). 

\appendix
%====================================================
\section{Field theoretical formalism of atom interferometer}
%====================================================
Usually, the light-atom interaction happens in a $\Lambda$-level system with lower energy levels $|1\rangle,|2\rangle$ which are the hyperfine structures and a higher energy level $|3\rangle$, as shown in Fig.\,\ref{fig:interaction}.  The atomic system is interacting with optical field which is off-resonant with respect to the energy gap of $|1\rangle-|3\rangle$ and $|2\rangle-|3\rangle$. The dynamics of this system can be reduced to an effective $|1\rangle-|2\rangle$ dynamics by adiabatically eliminating the energy level $|3\rangle$. Moreover, the atom states $|1\rangle$ and $|2\rangle$ are also associated with different linear center of mass momentum. The non-relativistic Hamiltonian of such a system, in a single-particle formalism, can be written as:
\be
\begin{split}
\hat H=-\frac{\nabla_i^2}{2m}+\sum_i\hbar\omega_i|i\rangle\langle i|&+E_c(x) d_{13}|1\rangle\langle 3|+E_p(x) d_{23}|2\rangle\langle 3|\\
&+h.c,
\end{split}
\ee
where $m$ is the inertia mass of the atom with center of mass momentum $p_i=-i\hbar\nabla_i$, $\omega_i$ corresponds to the energy of the internal electron state and $d_{ij}$ describes the dipole moment of the electron and it is convenient to define them as real numbers. The summation is over all three energy levels. Since the $|1\rangle-|2\rangle$ transition is forbidden, therefore $d_{12}=0$.  For extension to the multi-particle system and furthermore establishing a field-theoretique approach, we need to do \emph{second quantisation}\,\cite{Landauvol3}.\\

\subsection{Free fields}

For doing second quantisation, we need to first distinct the group of non-interacting identical particles, or equivalently, free theory. Tracing out the internal energy levels for the non-interaction single particle Hamiltonian, we have:
\be
\hat H_i=\hbar\omega_i-\frac{\nabla_i^2}{2m},
\ee
Assuming that there are $N-$ identical particles in group $i$, we have a multi-particle Hamiltonian as:
\be
\mathcal{\hat H}_i=\sum^N_a \left[\hbar \omega_i-\frac{\nabla_{ai}^2}{2m}\right].
\ee
where $a$ is the index of particles, and $\nabla_{ai}$ only act on the coordinate of $a$-th particle belong to group $i$.

Following the standard method of doing second quantisation, and choosing the orthnormal eigenfunctions of particles to be their momentum eigenstates, we have:
\be
\mathcal{\hat H}_i=\int \frac{d^3p}{(2\pi)^3}\left(\hbar\omega_i+m+\frac{p^2}{2m}\right)\hat a_p^\dag\hat a_p,
\ee
where the rest mass term is also included here. Apparently, this is a non-relativistic Hamiltonia. The full relativistic form is:
\be
\mathcal{\hat H}_i=\int \frac{d^3p}{(2\pi)^3}\left(\sqrt{m_i^2+p^2}\right)\hat a_p^\dag\hat a_p,
\ee
where $m_i=m+\hbar\omega_i$ with the physical meaning that the internal electron-nuclei interaction energy also contributes to the ``rest mass" of the atom. It is clear that such a Hamiltonian can be derived from a canonically quantised Klein-Gordon field with action:
\be
S_i=\int d^4x\left(\frac{1}{2}\partial_\mu\phi_i\partial^\mu\phi_i+m_i^2\phi_i^2\right).
\ee
Since we are now discussing a one-dimensional mode of atom interferometers, we will apply the para-axial approximation in order to reduce the above 3+1 action to a 1+1 action by integrating out the transversal components.

The relationship between 1+1 fields and the 3+1 fields is given as follows.
The plane wave expansion of the fields is:
\be
\phi_{3+1}(x)=\int \frac{dk_z}{(2\pi)(2\omega)}e^{-i\omega t+ik_z z}\int \frac{d^2k_\perp}{(2\pi)^2} e^{i\vec{k}_\perp\vec{x}_\perp}\hat a_k+c.c.
\ee
We thus have the dimensional relation: $[a_k]=[{\rm m}]^{3/2}[{\rm Hz}]^{1/2}$. While for 1+1 fields, we have:
\be
\phi_{1+1}(x)=\int \frac{dk_z}{(2\pi)(2\omega)}e^{-i\omega t+ik_z z}a_{k_z}+c.c,
\ee
where $[a_{k_z}]=[{\rm m}]^{1/2}[{\rm Hz}]^{1/2}$.
Denoting the cross sectional area of field beams as $A$, we then integrate out the transversal components and define:
\be
a_{k_z}\equiv\int \frac{d^2k_\perp}{(2\pi)^2}a_k e^{i\vec{k}_\perp\vec{x}_\perp}\sqrt{A},
\ee
so that 
\be
\begin{split}
\phi_{3+1}(x)&=\int \frac{dk_z}{(2\pi)(2\omega)\sqrt{A}}e^{-i\omega t+ik_z z} a_{k_z}+c.c\\
&=\frac{1}{\sqrt{A}}\phi_{1+1}(x).
\end{split}
\ee

The 3+1 form for electromagnetic field can be written as:
\be
\hat E_{c/p}=\int\frac{d^3k}{(2\pi)^3}\sqrt{\frac{\hbar\omega_k}{\epsilon_0}}\hat a^{c/p}_ke^{-i(\omega_k t-kx)}+{\rm h.c},
\ee
and following the sam eapproach, one can show that it can be effectively described by a 1+1 scalar field under paraxial approximation:
\be
E_{c/p}(z,t)=\sqrt{2\omega_{c/p}}\sqrt{\frac{\hbar\omega_{c/p}}{\epsilon_0A}}\phi_{c/p}(z,t).
\ee

\subsection{Interaction fields}
In a similar way, we can do second quantisation to the interaction Hamiltonian as follows:
\be
\hat H^{ij}_{p'p}=\int d^3x_a
\psi^*_{p'}(x_a)e^{iE^j_{p'}t}\hat E_{c/p}(x_a)d_{ij}\psi_p(x_a)e^{-iE^i_pt},
\ee
where $E^i_p=m_i+p^2/(2m_i)$, and $\psi_p(x_a)$ is the momentum eigenfunctions of atoms in the coordinate representation. Using box normalisation condition, we have $\psi_p(x_a)={\rm exp}(-i\vec{p}\vec{x}_a)/\sqrt{V_a}$. Substituting the second quantisation form of the electromagnetic field:
\be
\hat E_{c/p}=\int\frac{d^3k}{(2\pi)^3}\sqrt{\frac{\hbar\omega_k}{\epsilon_0}}\hat a^{c/p}_ke^{-i(\omega_k t-kx)}+{\rm h.c},
\ee
leads to:
\be
\begin{split}
\hat H^{ij}_{p'p}=&d_{ij}\hat a^{j\dag}_{p'}\hat a^i_p\int \frac{d^3x_ad^3k}{(2\pi)^3}
\psi^*_{p'}(x_a)\psi_p(x_a)e^{-ikz_a}e^{i(E_{p'}-E_p-\omega_k)t}\hat a^{c/p}_k\\
&+(\hat a_k\rightarrow \hat a_k^\dag, k\rightarrow-k, \omega\rightarrow-\omega)+{\rm h.c}.
\end{split}
\ee
Performing the integral on $x_a$, we have:
\be
\begin{split}
\hat H^{ij}_{p'p}=&\sqrt{\frac{\hbar\omega_0}{\epsilon_0V_a^2}}\frac{d_{ij}}{(2\pi)^3}e^{i(E^j_{p'}-E^i_p-\omega_{p'-p})t} \hat a^{j\dag}_{p'}\hat a^i_p\hat a^{c/p}_{p-p'}\\
&+(\text{non rotating wave terms})+{\rm h.c}.
\end{split}
\ee
The energy difference in the exponential can be written as:
\be
m_i+\frac{p^2}{2m_i}+\omega_{c/p}-\left(m_j+\frac{(p+k_c)^2}{2m_j}\right),
\ee
which clearly describes the energy transfer of the photon-absorption process. In practice,  the non rotating-wave terms can be safely ignored. The total interaction Hamiltonian will be:
\be
H=\sum_{p,p'} H_{p,p'},
\ee
where $p,p'$ are discretised momenta. In the continuous case, we have:
\be
\begin{split}
H=&\int \frac{d^3p}{(2\pi)^3} \frac{d^3p'}{(2\pi)^3}\sqrt{\frac{\hbar\omega_0}{\epsilon_0}}
\frac{d_{ij}}{(2\pi)^3}\hat a^{j\dag}_{p'}\hat a^i_p\hat a^{c/p}_{p-p'}e^{i(E^j_{p'}-E^i_p-\omega_{p'-p})t},\\
&+{\rm h.c}
\end{split}
\ee
where for the atomic operators: $\hat a_p\rightarrow\sqrt{V_a}\hat a_p$ (note that the dimension changes correspondingly).

This Hamiltonian can be derived from the following interaction action:
\be
S^{\rm full}_{\rm int}=g\int d^4x \phi_A(x)\phi_3(x)\phi_{c/p}(x),
\ee
where:
\be
g=\sqrt{\frac{\hbar\omega_0}{\epsilon_0}}\frac{d_{ij}}{(2\pi)^9}(2\omega_A)^{1/2}(2\omega_c)^{1/2}(2\omega_{c/p})^{1/2}.
\ee

Reducing it to 1+1 D case, the $g$ will change as:
\be
g=\sqrt{\frac{\hbar\omega_0}{\epsilon_0A}}\frac{d_{ij}}{(2\pi)^9}(2\omega_A)^{1/2}(2\omega_c)^{1/2}(2\omega_{c/p})^{1/2},
\ee
and we have an 1+1 interaction action:
\be\label{eq:interactionaction}
S=g_{13}\int dzdt \phi_A\phi_3\phi_c+g_{23}\int dzdt\phi_B\phi_3\phi_p.
\ee

\subsection{Effective interaction}
The above full interaction Hamiltonian can be further reduced to an effective Hamiltonian describing the transition between energy levels $\ket{1}-\ket{2}$ by integrate out the $\phi_3(x)$ field, which can be done solving Heisenberg equation of motions for the field $\hat \phi_3(x)$ (under canonical quantisation).

The Heisenberg equation of motion for $\hat \phi_3 (x)$corresponds to the above action is:
\be
(\Box+m_3^2)\hat \phi_3=g_{13}\hat \phi_A\hat \phi_c+g_{23}\hat \phi_B\hat \phi_p,
\ee
Written in the rotating frame of $\omega_{30}$ and $k_{30}$ (they are the Compton frequency and Compton wave vector for the atoms on level $|3\rangle$), we have:
\be
\hat\phi_3(x)=\tilde\phi_3(x)e^{-i\omega_{30}t+ik_{30}z}+{\rm h.c},
\ee
and the equations of motion for $\tilde{\phi}_3$ is given approximately by:
\be
\begin{split}
(\partial_t+v\partial_z)\tilde\phi_3(x)&=\frac{ig_{13}}{2\omega_{30}}\tilde \phi_c\tilde \phi_A e^{i\Delta\omega_{3A}t-i\Delta k_{3A}z}\\
&+\frac{ig_{23}}{2\omega_{30}}\tilde \phi_p\tilde \phi_B e^{i\Delta\omega_{3B}t-i\Delta k_{3B}z},
\end{split}
\ee
where we have already ignored the non-rotating-wave terms, which oscillates at frequencies $\omega_{30}\pm\omega_{c/p0}\mp\omega_{A/B0}$ or $\omega_{30}+\omega_{c/p0}+\omega_{A/B0}$. The $\Delta\omega_{3A/B}, \Delta k_{3A/B}$ are defined as:
\be
\begin{split}
\Delta\omega_{3A/B}&=\omega_{30}-\omega_{c/p}-\omega_{A/B0}\\
&\approx\omega_{3A/3B}-\omega_{c/p0}+v_{A/B}(k_{30}-k_{A/B0}),\\
\quad\Delta k_{3A/B}&=k_{30}-k_{c/p}-k_{A/B0},
\end{split}
\ee
where $\omega_{3A/B}$ are the energy level difference between $\ket{3}$ and $\ket{1},\ket{2}$. In ideal case, we assume that the phase matching is satisfied $\Delta k_{3A/B}=0$. We also define $\Delta\omega_0:=\omega_{3A}-\omega_c=\omega_{3B}-\omega_p$ and if we futher choose $\tilde\phi_3$ to work iin the rotating frame of $\Delta \omega_0$, the equation becomes:
\be
\begin{split}
\Delta\omega_0\bar{\phi}_3(x)+(\partial_t+v\partial_z)\bar\phi_3(x)&=\frac{g_{13}}{2\omega_{30}}\tilde \phi_c\tilde \phi_A e^{iv_Ak_ct}\\
&+\frac{g_{23}}{2\omega_{30}}\tilde \phi_p\tilde \phi_B e^{iv_Bk_pt},
\end{split}
\ee
Under the adiabatic approximation that \emph{the typical time scale of varying $\bar\phi_3(x)$ is much smaller than $1/\Delta\omega_0$}, the dynamics of $\bar\phi_3(x)$ is slaved by the right hand side of the above equation:
\be\label{eq:phi3}
\bar\phi_3(x)\approx\frac{g_{13}}{2\Delta\omega_0\omega_{30}}\tilde \phi_c\tilde \phi_A e^{iv_Ak_ct}+\frac{g_{23}}{2\Delta\omega_0\omega_{30}}\tilde \phi_p\tilde \phi_B e^{iv_Bk_pt},
\ee
while the positive frequency branch of $\hat \phi_3(x)$ can be recovered by changing to the original frame: $\bar\phi_3(x){\rm exp}[i\Delta\omega_0t-i\omega_{30}t+ik_{30}z]$. Substituting $\hat\phi_3(x)$ back into the interaction Hamiltonian Eq.\,\eqref{eq:interactionaction}, we have:
\be
\begin{split}
&\mathcal{\hat H}_{\rm int}=\mathcal{\hat H}^{\rm Stark}_{\rm int}+\mathcal{\hat H}^{\rm AB}_{\rm int},\\
&\mathcal{\hat H}^{\rm Stark}_{\rm int}=\frac{g^2_{13}}{2\Delta\omega_0\omega_{30}}\int dz\hat \phi_A(x)\underline{\hat \phi_c(x)
 \hat \phi_c(x)}\hat \phi_A(x)\\
 &\qquad\quad+\frac{g^2_{23}}{2\Delta\omega_0\omega_{30}}\int dz \hat \phi_B(x)\underline{\hat \phi_p(x)
 \hat \phi_p(x)}\hat \phi_B(x),\\
 &\mathcal{\hat H}^{\rm Raman}_{\rm int}=2\times\frac{g_{13}g_{23}}{2\Delta\omega_0\omega_{30}}\int dz\hat \phi_A(x)\hat \phi_c(x) \hat \phi_p(x)\hat\phi_B(x).
\end{split}
\ee
Here, the $\mathcal{\hat H}^{\rm Stark}_{\rm int}$ corresponds to the AC Stark shift of the mass of the atom fields, which is typically very small compare to the internal energy and rest mass thereby negligible\,\footnote{The mass correction due to AC Stark shift is position dependent here, which is not surprising since the light-atom interaction happens in a local region of space-time.}.The  $\mathcal{\hat H}^{\rm Raman}_{\rm int}$ term is the one we are interested in, i.e.\, describing the Raman transition between $\ket{1}$ and $\ket{2}$ induced by control and passive fields. This term can also be correspondence to an effective four-scalar field interaction action:
\be
S_{\rm int}=g\int d^2x \phi_{A}(x)\phi_{B}(x)\phi_{c}(x)\phi_{p}(x),
\ee
where $g$ represents the coupling coefficient in $\mathcal{\hat H}^{\rm Raman}_{\rm int}$.\\

%====================================================
\section{Field Quantisation}
%====================================================
\subsection{Definition of field operators}
As discussed above, The one-dimensional light-atom interaction model can be described by the following interaction action as:
\begin{equation}\label{eq:action}
S_{\rm int}=g\int d^2x \phi_{A}(x)\phi_{B}(x)\phi_{c}(x)\phi_{p}(x)
\end{equation}
where $x=(t,z)$. These free scalar fields, after \emph{canonical quantisation} can be expanded as:
\be\label{eq:free-expansion}
\hat\phi^{(0)}_j(x)=\int\frac{dk_j}{2\pi 2\omega_j}\left[\hat a(k_j)e^{-i\omega_j (t-t_{j0})+ik_j (z-z_{j0})}+{\rm h.c}\right],
\ee
where the $\hat a(k_j)$ is an annihilation operator which is covariant under Lorentz transformation and $j=A,B,c,p$. The $\omega_j$ and $k_j$ are related by dispersion relation $\omega_j^2=k_j^2+m_j^2$, while for optical fields $m_j=0$. The $t_{j0},z_{j0}$ determine the phase reference point.\\

When we introduce the interaction, the Heisenberg operators will be modified, according to:
\be
\hat O(t,z)=\hat U^\dag_I(-\infty,t)\hat O^{(0)}(t,z)\hat U_I(-\infty,0),
\ee
where $\hat O^{(0)}(t,z)$ and $\hat O(t,z)$ are the operators whose evolution are governed by the free and full Hamiltonian, respectively. The $\hat U_I(t_1,t_2)$ is the evolution operator in the interaction picture. Therefore, in the interaction case, one can have a full field operator given by:
\be
\hat\phi_j(x)=\int\frac{dk_j}{2\pi 2\omega_j}\left[\hat A(k_j,t)e^{-i\omega_j (t-t_{j0})+ik_j (z-z_{j0})}+{\rm h.c}\right],
\ee
where $\hat A(k_j,t)=\hat U_I^\dag(0,t)\hat A(k_j,0)\hat U_I^\dag(0,t)$.\\

Since all these fields have a WKB trajectory in real experiment, they can take a scale-separated form as:
\be\label{eq:sva}
\hat{\phi}_{j}(x)=\tilde{\phi}_{j}(x)e^{-i \omega_{j0}(t-t_{j0})+ik_{j0}(z-z_{j0})}+{\rm h.c}
\ee
where the exponents describes the fast-oscillating part of the field and the tilde operators describe the slowly varying amplitudes. Using the above definition, one can obtain the relation between $\hat A-$ operator and $\tilde{\phi}$ as:
\be\label{eq:phiArelation}
\begin{split}
\tilde{\phi}_{j}(t,z)=\int \frac{dk_{\rm A}}{2\pi 2\omega_{j}}[&\hat{A}(t,k_{j})e^{-i(\omega_{j}-\omega_{j0})(t-t_{j0})+i(k_{j}-k_{j0})(z-z_{j0})}\\&+{\rm h.c.}]
\end{split}
\ee
Also note that the transformation $\hat U(0,t)$, as an unitary transformation, will not affect the commutation relation, therefore we have:
\be
[\hat A_j(t, k),\hat A^\dag_j(t,k')]=2\omega_j\delta(k-k').
\ee
It is worth to note that in the spatial domain, the $\hat A$ and the $\tilde \phi$ is related by (take A-field as an example):
\be
\tilde\phi^{(+)}(x)=\hat A(t,z-z_A^t)/(2\omega_{A0}),
\ee
where $z-z^t_A$ ($z^t_A=z_A(t_0)+v_A(t-t_0)$) comes from the propagation and the first $t-$argument comes from the perturbation due to the 4-scalar interaction. 

\subsection{Quantum states of fields}
In the experimental setup, the control/passive field can be well approximated to have a rectangular profile, and the states of, e.g. the control field is given by:
\begin{equation}\label{eq: controlstate}
\ket{\psi_{c}}={\rm exp}\left[\int \frac{dk_{c}}{2\omega_c}\left(\alpha_{c}(k_{c})\hat{a}_c^\dagger(k_{c})-{\rm h.c.}\right)\right]\ket{0}_c,
\end{equation} 
where
\be
\alpha_c(k_c)=\bar \alpha_c\frac{\sin{\pi a\delta k}}{\sqrt{a}\pi\delta k},\quad \delta k=k_c-k_{c0},
\ee
where $2\pi a$ is the width of this rectangular wave and $\bar \alpha_c$ is the coherent amplitude.

It is easy to show the $\phi-$ waveform
\be
\begin{split}
\bar \phi^{(+)}=&\int \frac{dk}{2\pi}\left(\frac{1}{2\omega}\right)\bar \alpha_c
\frac{\sin{\pi a\delta k}}{\sqrt{a}\pi\delta k}e^{i(k_0+\delta k)(x-t-x_0+t_0)}\\
&\approx\frac{\bar \alpha_c}{2\omega_0}{\rm Rect}_a[x-t-x_0+t_0]e^{ik_0(x-t-x_0+t_0)}.
\end{split}
\ee

Since the typical atom width is much smaller than the light pulse width, it can be approximated as almost a plane wave in the atom-light interaction region by ${\rm lim}_{a\rightarrow \infty}a\times{\rm sinc}(\pi ax)=\delta (x)$.

The initial state of atomic cloud is a Gaussian profile, given as:
\be
|\psi_{A}\rangle={\rm exp}\left[\int \frac{dk_{A}}{2\omega_A}\left(\alpha_{A}(k_{A})\hat{a}_A^\dagger(k_{A})-{\rm h.c.}\right)\right]\ket{0}_A,
\ee
with
\be
\alpha_{A}(k_{A})=\frac{\bar \alpha_A}{(2\pi)^{1/4}\Delta^{1/2}_A}{\rm exp}\left[-\frac{(k_A-k_{A0})^2}{4\Delta_A^2}\right],
\ee
while the $B$-field is a vacuum initially. Here the $\bar\alpha_A$ is the atom coherent amplitude, $\Delta_A$ is its width in the $k$-domain.

\subsection{Equations of motion: structures}
Following the standard canonical quantisation scheme, we have the Heisenberg equations of motion for the atomic and optical field as:
\be\label{eq:fieldeom}
\begin{split}
&(\Box+m_A^2)\hat{\phi}_{\rm A}=g\hat{\phi}_{B}\hat{\phi}_{c}\hat{\phi}_{p},\quad \Box\hat \phi_c=g\hat{\phi}_{A}\hat{\phi}_{B}\hat{\phi}_{\rm p}\\
&(\Box+m_B^2)\hat{\phi}_{\rm B}=g\hat{\phi}_{A}\hat{\phi}_{c}\hat{\phi}_{p}
,\quad\Box\hat\phi_p=g\hat{\phi}_{A}\hat{\phi}_{B}\hat{\phi}_{\rm c},
\end{split}
\ee
where $\Box=\partial^2_t-\nabla^2$ (in 1-dimensional case here, $\Box=\partial^2_t-\partial_z^2$).
These equations have the same form as their classical counterparts, this is because different $\hat \phi_j$ belongs to different Hilbert space and their operators are commute with each other. Substituting Eq.\,\eqref{eq:sva}, we have the approximated equations for the slowly varying operator's positive frequency part (we take $c=1$ and their Hermitian conjugates are ignored for brevity):
\be
\begin{split}
&(\partial_t+v_{A}\partial_z)\tilde\phi^{+}_{A}=g_A\tilde\phi^+_{B}\tilde\phi^+_p\tilde\phi^-_c,\\
&(\partial_t+v_{B}\partial_z)\tilde\phi^{+}_{B}=g_B\tilde\phi^+_{A}\tilde\phi^-_p\tilde\phi^+_c,\\
&(\partial_t+\partial_z)\tilde\phi^+_{c}=g_c\tilde\phi_A^-\tilde\phi_B^+\tilde\phi^-_{p},\\
&(\partial_t-\partial_z)\tilde\phi^+_{p}=g_p\tilde\phi_A^+\tilde\phi_B^-\tilde\phi^+_{c}.
\end{split}
\ee
Here, $v_{A/B}$ the WKB velocity of atom wave packet A/B and the two optical fields are propagating along the opposite directions. The coupling constants are defined as: $g_j=ig/(2\omega_{j0})$, where $j=A,B,c,p$. In deriving the above equations of motion, we take the leading non-relativistic approximation so that $v_{A/B}\approx k_{A/B}/m_{A/B}$ and we also use the rotating wave approximation, namely, we only keep those terms satisfying:
\be\label{eq:rwa}
k_{p0}-k_{c0}=k_{A0}-k_{B0},\quad\omega_{c0}-\omega_{p0}=\omega_{B0}-\omega_{A0}.
\ee
These conditions, under non-relativistic approximation and $m_B\approx m_A$, has the clear physical meaning of relativistic Doppler effect:
\be
\omega_{c0}-\omega_{p0}\approx v(k_{A0}-k_{B0}),
\ee
where $v\approx v_A\approx v_B$ is the approximate speed of atom. The Hermitian conjugation of these equations can also be easily obtained. In the following, we are going to solve these equation in a perturbative way.\\

\section{Mean field solutions}
Typically, in an interferometric process, the light-atom interaction time is very short compare to the free evolution time of the atom cloud, and the centre of mass velocity of the atom cloud is very low, typically $\sim 2$\,cm/s. Therefore to the leading order, we can treat the atom centre of mass motion to be static during the interaction process, that is, $v_A\approx v_B\approx 0$. We also note that the spatial size of optical fields are much larger than the size of the atom cloud, therefore we can approximate the mean value of the optical fields to be almost constants during the interaction process.

The zeroth-order of the equations of motion is simple:
\be
\begin{split}
&\partial_t\bar\phi^{+}_{A}=(g_A\bar\phi^+_p\bar\phi^-_c)\bar\phi^+_{B},\quad (\partial_t+\partial_z)\bar\phi^+_{c}=0,\\
&\partial_t\bar\phi^{+}_{B}=(g_B\bar\phi^-_p\bar\phi^+_c)\bar\phi^+_{A},\quad (\partial_t-\partial_z)\bar\phi^+_{p}=0.\\
\end{split}
\ee
We ignore the r.h.s. of the equation for the optical field because the photon number is much larger than the atom number. Since $\bar \phi_p$ and $\bar \phi_c$ are almost constant, therefore we can rewrite the zeroth-order atom equations to be:
\be
\partial_t\bar\phi^{+}_{A}=-i\Omega e^{i\varphi_{pc}}\bar\phi^+_{B},\quad
\partial_t\bar\phi^{+}_{B}=-i\Omega e^{-i\varphi_{pc}}\bar\phi^+_{A},
\ee
where $\varphi_{pc}$ is the phase difference between the control field and passive field, $\Omega$ is the Ramsey frequency. Here, we make use of the approximation $g_A=g_B:=g_a$ thereby $\Omega=|g_a\bar \phi_p\bar\phi_c|$. The full solution of this equation is:
\be
\begin{split}
&\bar\phi_A^+(t)=\bar\phi^+_A(0)\cos{\Omega t}-i\bar\phi^+_B(0)e^{i\varphi_{cp}}\sin\Omega t,\\
&\bar\phi_B^+(t)=\bar\phi^+_B(0)\cos{\Omega t}-i\bar\phi^+_A(0)e^{-i\varphi_{cp}}\sin\Omega t.
\end{split}
\ee
The gravitational wave signal will be carried by the $\varphi_{cp}$. Clearly, this is where the matrix $\mathbb{M}(\theta,\varphi)$ in the main text comes from.

\bibliography{atom_interferometry}

\end{document}